\documentclass[pdflatex,sn-mathphys-num,iicol]{sn-jnl}

\usepackage{graphicx}%
\usepackage{multirow}%
\usepackage{amsmath,amssymb,amsfonts}%
\usepackage{amsthm}%
\usepackage{mathrsfs}%
\usepackage[title]{appendix}%
\usepackage{xcolor}%
\usepackage{textcomp}%
\usepackage{manyfoot}%
\usepackage{booktabs}%
\usepackage{algorithm}%
\usepackage{algorithmicx}%
\usepackage{algpseudocode}%
\usepackage{listings}%

\theoremstyle{thmstyletwo}%

\theoremstyle{thmstylethree}%

\begin{document}

\title[Article Title]{Cluster emission and its impact on the r-process nucleosynthesis}

\author*[1,2]{\fnm{Natalia} \sur{Thomas}}\email{natalia.thomas@ulb.be} 

\author[1,3,4]{\fnm{Silvia} \sur{Bara}} 

\author[1]{\fnm{Thomas Elias} \sur{Cocolios}} 

\author[2]{\fnm{Stephane} \sur{Goriely}} 

\author[5]{\fnm{Boris Andel}} 

\author[6]{\fnm{Andrei N.} \sur{Andreyev}} 

\author[7,1]{\fnm{Alberto} \sur{Camaiani}} 

\author[6,8]{\fnm{James G.} \sur{Cubiss}} 

\author[1]{\fnm{Hilde} \sur{De Witte}} 

\author[9]{\fnm{Zoe} \sur{Favier}} 

\author[1,10]{\fnm{Michael} \sur{Heines}} 

\author[1,11]{\fnm{Fedor} \sur{Ivandikov}} 

\author[1]{\fnm{Jake D.} \sur{Johnson}} 

\author[1]{\fnm{Jozef} \sur{Klimo}} 

\author[12]{\fnm{Razvan} \sur{Lica}} 

\author[5]{\fnm{Jozef} \sur{Mi$\check{\text{s}}$t}} 

\author[6]{\fnm{Chris} \sur{Page}} 

\author[1]{\fnm{Riccardo} \sur{Raabe}} 

\author[5]{\fnm{Adam} \sur{Sitar$\check{\text{c}}$\'{\i}k}} 

\author[1]{\fnm{Viktor} \sur{Van Den Bergh}} 

\author[1]{\fnm{Piet} \sur{Van Duppen}} 

\author[6]{\fnm{Zixuan} \sur{Yue}} 

\author[1]{\fnm{Ahmed} \sur{Youssef}} 

\affil[1]{\orgname{KU Leuven}, \orgdiv{Instituut voor Kern- en Stralingsfysica}, \city{Leuven}, \postcode{3001}, \country{Belgium}}

\affil*[2]{\orgdiv{Institute d'Astronomie et d'Astrophysique}, \orgname{Université Libre de Bruxelles}, \postcode{1050}, \city{Brussels}, \country{Belgium}}

\affil[3]{\orgdiv{Université de Caen Normandie, ENSICAEN}, \orgname{LPC Caen}, \postcode{14000}, \city{Caen}, \country{France}}

\affil[4]{\orgdiv{Grand Accélérateur National d'Ions Lourds (GANIL)}, \orgname{CEA/DRF-CNRS/IN2P3}, \postcode{F-14076}, \city{Caen}, \country{France}}

\affil[5]{\orgdiv{Department of Nuclear Physics and Biophysics}, \orgname{Comenius University in Bratislava}, \postcode{842 48}, \city{Bratislava}, \country{Slovakia}}

\affil[6]{\orgdiv{School of Physics, Engineering and Technology}, \orgname{The University of York} \postcode{YO10 5DD}, \city{York}, \country{United Kingdom}}

\affil[7]{\orgdiv{INFN}, \orgname{Sezione di Firenze}, \postcode{50019}, \city{Firenze}, \country{Italy}}

\affil[8]{\orgdiv{School of Physics and Astronomy}, \orgname{The University of Edinburgh}, \postcode{EH9 3FD}, \city{Edinburgh}, \country{United Kingdom}}

\affil[9]{\orgname{CERN}, \postcode{1211}, \city{Geneva}, \country{Switzerland}}

\affil[10]{\orgname{PSI Center for Neutron and Muon Sciences}, \postcode{5232}, \city{Villigen}, \country{Switzerland}}

\affil[11]{\orgdiv{Accelerator Laboratory, Department of Physics} \orgname{University of Jyväskylä}, \postcode{40014}, \city{Jyväskylä}, \country{Finland}}

\affil[12]{\orgname{Horia Hulubei National Institute for R\&D in Physics and Nuclear Engineering}, \postcode{R-077125}, \city{Bucharest}, \country{Romania}}

\abstract{\unboldmath Cluster emission is an exotic decay mode between $\alpha$ decay and fission, in which a parent nucleus emits a cluster of nucleons heavier than an $\alpha$ particle, but lighter than what is usually considered a fission fragment. The properties of cluster emission were investigated by analyzing five high-energy events detected in a spectrum of a mass $A$ = 230 beam produced at ISOLDE (CERN). Under the assumption that these events were caused by cluster emission, the most likely parent-cluster pair responsible for the five high-energy events, was found to be $^{230}$Ra emitting $^{22}$O, with a branching ratio of ($4.3 \pm 1.9)\times \mathrm{10}^{-9}$. Four analytical formulas were used to estimate the partial half-lives of cluster emission for a group of neutron-rich nuclei. The decay rate was calculated for a selection of cluster nuclei for each parent isotope. The rates of all decay channels of cluster emission per parent nucleus were then included in calculations of the r-process nucleosynthesis in a neutron star merger in order to study the possible impact of cluster emission on the r-process nuclear production. The resulting isotopic abundance distributions were compared to those calculated for a case in which cluster emission was not considered. It was found that the inclusion of cluster emission decay rates from the simple analytical formulas available nowadays does not influence the results of the r-process nucleosynthesis.}


\maketitle

\section{Introduction}\label{sec1}
The \textit{rapid neutron capture}, or r-process, is a nucleosynthetic process responsible for the production of approximately half of all nuclei heavier than iron~\cite{1957RvMP...29..547B, 1965ApJS...11..121S, 1999A&A...342..881G}. During this process, seed nuclei rapidly absorb neutrons in chains before having a chance to decay, leading to the creation of exotic, neutron-rich isotopes. The point of equilibrium between neutron capture and photodisintegration reactions dictates the end of the chain. Subsequent $\beta^{-}$ decay converts neutrons into protons, synthesizing heavier elements. Thus, the r-process is a complex interplay of neutron-captures, photodisintegration reactions, and $\beta^{-}$ decays~\cite{ARNOULD200797, ARNOULD2020103766, KAJINO2019109}. Additionally, other decay processes are also at play, including various modes of fission of neutron-rich actinides, such as spontaneous, \mbox{$\beta$-delayed}, and neutron-induced fission. 
These processes can influence the nuclei production via the fission recycling process, in which heavy nuclei undergo a form of fission, thus ending the r-process path and simultaneously feeding the seed nuclei upon which the process can restart~\cite{2005NuPhA.758..587G, RevModPhys.93.015002}. As such, the careful consideration of all decay processes and nuclear reactions of relevance to the isotopes at play is of utmost importance to understand the heavy element production by the r-process.

Extensive research has been dedicated to the astrophysical site of the r-process nucleosynthesis. For the neutron capture chains of the r-process to occur, it must necessarily take place in highly energetic environments, which include extremely high temperatures ($T\gtrsim 10^9$ K) and neutron densities exceeding $10^{24}$ cm$^{-3}$~\cite{1957RvMP...29..547B, ARNOULD200797, ARNOULD2020103766, RevModPhys.93.015002}. These requirements cannot be met in quiescent stellar evolutionary phases, meaning the r-process must necessarily take place in explosive environments. The r-process has been described in terms of a ``weak" and a ``strong" component, depending on the available initial neutron density at the astrophysical site~\cite{WANAJO2006676}. Only the latter is predicted to synthesize the heaviest elements, and thus believed to be the dominant component responsible for solar-system r-process abundances\mbox{~\cite{2011ApJ...726L..15W, Wanajo_2012, RevModPhys.93.015002}}.

Core-collapse supernovae of massive stars had long been considered a primary site of the r-process through the presence of neutrino-driven winds~\cite{Woosley1994}, but so far no model of such an event has indicated the presence of a successful r-process, including those with a sophisticated neutrino transport treatment. At most, a ``weak" r-process has been predicted for this r-process site~\cite{PhysRevLett.109.251104, PhysRevC.86.065803, Wang_2023}. An alternative proposed site of the r-process is that of neutron star (NS) mergers. The conditions during an NS merger have long been envisioned to eject substantial amounts of matter~\cite{1974ApJ...192L.145L, 1976ApJ...210..549L, 1982ApL....22..143S}, allowing for the possibility of galactic enrichment with r-process nuclei. Subsequent hydrodynamical simulations of such an event, combined with nucleosynthesis calculations, predicted concrete mass ejections of $10^{-3}$ -- $10^{-2} \text{ solar mass }M_\odot $~\cite{1999A&A...341..499R, Goriely_2011, Wanajo_2014, 10.1093/mnras/stw2156}.

In 2017, the landmark gravitational wave observation GW170817~\cite{2017PhRvL.119p1101A} accompanied by an electromagnetic transient AT2017gfo~\cite{2017ApJ...848L..12A}, provided the first observational indications of the r-process during an NS merger. The analysis of the associated kilonova light curve and spectra allowed for insight into the produced r-process nuclei, whose decay powers these observables~\cite{2017Natur.551...80K}, with the identification of Sr in the GW170817 ejecta confirming NS mergers as an r-process site~\cite{2019Natur.574..497W, 2022MNRAS.515..631G}.

\textit{Cluster emission} is an exotic decay mode observed in heavy and superheavy nuclei, described as an intermediate process between $\alpha$ decay and fission~\cite{osti_6189038, rose_new_1984, Poenaru_1984}. The parent isotope decays through emission of a nucleus heavier than an $\alpha$ particle, but lighter than fission fragments. The authors of the establishing prediction~\cite{osti_6189038} theorized that the effects of shell closure generate additional fusion and fission reaction valleys~\cite{JAIN2023122597}. The subsequent experimental observations of cluster emission demonstrated the importance of shell effects to the process. Emission of clusters ranging from $^{14}$C to $^{34}$Si has been measured experimentally (See Ref.~\cite{bonetti_guglielmetti_2007} for a review). In these decays, the daughter nuclei tend to be nearly spherical closed-shell nuclei, with the stable doubly magic nucleus $^{208}$Pb ($Z$ = 82, $N$ = 126) and its neighboring nuclei being the most prominent examples. Pairing effects are also of significant importance to cluster radioactivity, strongly influencing which clusters can be emitted and from which parent nucleus~\cite{ROYER2001182}. Accordingly, the majority of emitters are even-even nuclei~\cite{BONETTI199332}. 

Through experimental measurements of cluster emission, the decay mode was found to be primarily governed by barrier penetration~\cite{BONETTI199332}. The phenomenon can then be characterized by incorporating this barrier penetration in different descriptions. Efforts have been made to describe cluster emission by methods belonging to both decay modes of fission and $\alpha$-particle emission. One description similar to that of $\alpha$-particle emission exploits the Gamow-like barrier penetration. In this type of approach, the cluster is assumed to be preformed before tunneling through the potential barrier. Thus, semi-empirical formulas relating decay Q-value to $\alpha$-particle emission half-life can be adapted to describe cluster emission half-lives as well~\cite{PhysRevC.70.017301}.
Another approach considers the decay to be a case of very asymmetric fission~\cite{Poenaru_1984, PhysRevC.32.572}. A fission-like description was also employed in the original prediction of cluster emission~\cite{osti_6189038} and has guided successful experiments of the phenomenon~\cite{bonetti_guglielmetti_2007}. More recently, the description of cluster emission as very asymmetric fission has also been examined fully microscopically for individual nuclei~\cite{warda2011}.
Each of these descriptions have been successful in closely reproducing experimental cluster emission half-lives \cite{PhysRevC.70.017301, PhysRevC.32.572, bonetti_guglielmetti_2007}. In particular, the success of models that describe cluster emission as a highly asymmetric fission process, raises the question of whether it might affect the results of r-process nucleosynthesis in a way similar to fission. Thus, if cluster emission rates can compete with those of other decay modes for nuclei along the r-process path, it may impact the produced abundance distribution as well.

All current empirical evidence of cluster emission points to an exotic decay process, with branching ratios ranging from $10^{-4}$ to $10^{-16}$. While such branching ratios would imply a negligible impact by cluster emission on the r-process, so far no measurements are available for the heavy nuclei relevant to the r-process nucleosynthesis. 

A recent theoretical study of cluster emission in the superheavy region predicted for a series of isotopes of $Z$ = 118 and $Z$ = 120 half-lives of less than a second or even much lower, down to $10^{-18} \text{ s}$~\cite{SAXENA2024122867}. For these isotopes, partial decay rates for cluster emission are predicted to be higher than for other decay modes like $\beta$, $\alpha$, or fission decay. Ref.~\cite{SAXENA2024122867} reported on neutron-deficient nuclei; however, during r-process nucleosynthesis, nuclei present will instead be neutron-rich. Thus, insight into cluster emission half-lives must be extended to neutron-rich nuclei in order to estimate its effect on the abundances resulting from the r-process nucleosynthesis.

This work aims to investigate and quantify the possible effect of cluster emission on the r-process nucleosynthesis in NS mergers by estimating the rates of cluster emission for a group of heavy nuclei ($A < 200$) relevant to the r-process. Since r-process calculations require decay rates across a large region of the nuclear chart, for which the computational cost of the fission-like approach is prohibitive, we instead adopt $\alpha$-like semi-empirical formulas, validated above against the same experimental half-lives, to estimate the rates of cluster emission. Four semi-empirical formulas relating decay Q-value to partial half-life were used, and the resulting decay rates were included into the r-process nucleosynthesis post-processing calculations of a NS merger model. 

In Sect. \ref{clusteremission}, we report on the observation of five high-energy events in a spectrum of mass $A$ = 230 and study the possibility of these being cluster emission events. In Sect. \ref{rprocesstheory}, a selection of established semi-empirical formulas is introduced and used to predict the cluster emission half-lives for a network of nuclei relevant to the r-process. These half-lives are subsequently included in r-process nucleosynthesis calculations in the ejecta of a NS merger. Sect.~\ref{results} presents the results for these cluster emission half-lives and their effect on the r-process nucleosynthesis and a brief discussion. 

\section{High-energy events in a spectrum of $A = 230$}
\label{clusteremission}

\begin{figure}[]
\centering
\includegraphics[width=\columnwidth]{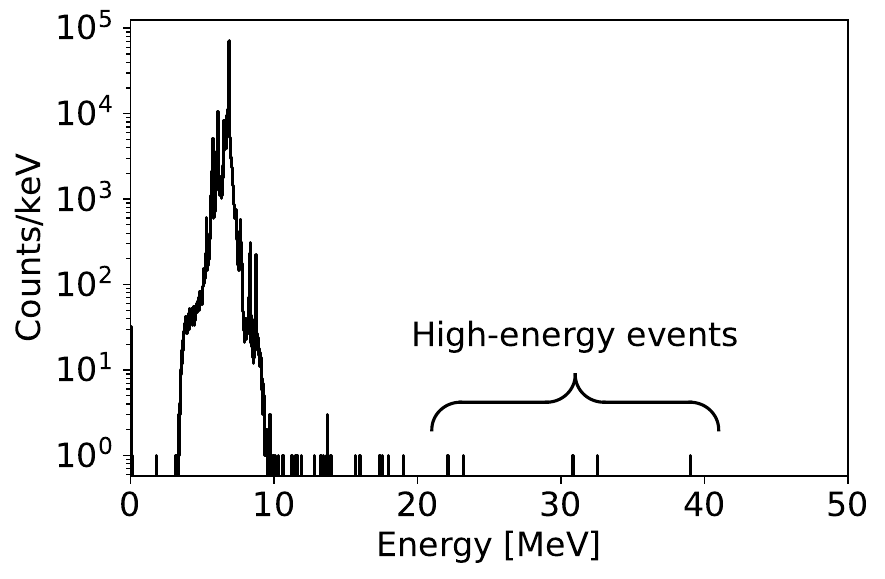}
\caption{The mass $A$ = 230 spectrum from an annular silicon detector featuring five high-energy events in the range of 20 to 40 MeV.}\label{230ra}
\end{figure}

Currently available experimental data on cluster emission are limited to a set of nuclei ranging from $^{221}$Fr to $^{242}$Cm~\cite{bonetti_guglielmetti_2007}. Therefore, each addition offers opportunities to benchmark established models. In the framework of an experimental campaign focused on measuring $\beta$-delayed fission of $^{230,232,234}$Ac at the ISOLDE facility at CERN, five high-energy events were identified in a silicon annular detector spectrum of a mass-separated radioactive ion beam with $A$ = 230. The data were collected over several runs, during which the beam was implanted for approximately 1 h and then allowed to decay for several hours, while continuously measuring. In total, approximately 47 h of data were recorded, including 5 h  of implantation and 42 h of decay. The description of the setup and experimental methods can be found in the corresponding paper of the campaign~\cite{PhysRevC.111.065803, lirias4273769, bara_2024_14358715}. For the spectrum of mass $A$ = 230 relevant here, the radioactive ion beam was a mixture of $^{230}$Fr and $^{230}$Ra $\beta^-$-decaying into $^{230}$Ac. In Fig. \ref{230ra}, five high-energy events are visible in a range of 20 to 40 MeV.

\subsection{Validity of the cluster emission assumption}

The possibility of the high-energy events originating from contamination of the beam, noise, \mbox{$\alpha$ summing}, or very low-energy fission fragments was investigated. Additionally, the calibration of the detector at high energies was analyzed to study the deviation of the energies of the five high-energy events measured. 

The detector was calibrated with $\alpha$ particles with energies ranging between 5.115(2) - 8.784(1) MeV. Thus, energy calibration can deviate substantially at higher energies. By taking the deviation from the experimental value of the $\alpha$-decay peaks \cite{KONDEV2025346, BROWNE20111115, KONDEV2021509, MARTIN20071583, MORSE2026409, KONDEV2011707, KONDEV2018382, SINGH2013661, SHAMSUZZOHABASUNIA2014561, BASUNIA2022475, CHEN2015373} to the measured energies and linearly extrapolating this to the energy range where the five events were found, the accuracy of the calibration at high energies could be estimated. The deviation was found to be no larger than 0.35 MeV using a 95\% confidence interval (CI). In order for the deviation to affect our conclusions it would need to be in the range of several MeV, as the events are already spread over a $\sim25$ MeV range. 

While both molecular contamination from $^{211}$Ra$^{19}$F and gaseous contamination from various isotopes of Rn were found, none of the beam contaminants are known cluster emitters, with the exception of $^{224}$Ra from the decay chain of $^{224}$Rn, and $^{230}$Th, a daughter of $^{230}$Ac~\cite{bonetti_guglielmetti_2007}. For $^{224}$Ra, the branching ratio for cluster emission compared to $\alpha$ decay is known to be $10^{11}$~\cite{224Ra}. Therefore, considering the statistics collected for its main $\alpha$ line at 5.6854(2) MeV~\cite{BROWNE20111115} ($10^4$ events), we can expect $10^{-7}$ cluster events in this measurement, making it highly unlikely that $^{224}$Ra is the source of the observed events. As for $^{230}$Th, this isotope will be produced through the $\beta^-$ decay of $^{230}$Ac. However, no $\alpha$-particle emission from $^{230}$Th is detected in the spectrum, in accordance with its long half life ($7.54 \times 10^4$ y~\cite{MORSE2024259}) compared to the implantation and measurement times. 

We can estimate an upper limit on the expected number of cluster emission events from $^{230}$Th by requiring that the expected number of counts remain below the order of magnitude ($10^{1}$ counts) of the continuum at the energy of its main $\alpha$ lines (4.6 MeV~\cite{MORSE2024259}), such that any contribution would be indistinguishable from the background. The cluster-to-$\alpha$ branching ratio of $\sim 10^{-13}$ then implies an upper limit of $10^{-12}$ expected cluster emission events from $^{230}$Th. The possibility that the observed events originated from contaminants of nearby masses was also investigated and ruled out in Ref.~\cite{PhysRevC.111.065803}.

Electronic noise cannot be definitively excluded as a cause, but we note that the mass \mbox{$A$ = 230} spectrum consists of different runs, during which disturbances were not experienced, with each run registering at least one count in the region between 20 to 40 MeV. Additionally, a spectrum of $A$ = 232, with a similar measurement time, taken during the same experimental campaign registered no event in this energy range. Thus, noise is not expected to be responsible for the five high-energy events.

Based on the count-rate of $\alpha$ decay in the \mbox{$A = 230$} spectrum, the possibility of $\alpha$ summing was examined. In order to reach energies of 20 to 30 MeV through $\alpha$ summing we examine the summing of either three high-energy \mbox{(7.5 - 10 MeV)} or four low-energy \mbox{(3 - 7.5 MeV)} $\alpha$ particles. For a total measurement time of \mbox{$t_\text{meas} = 1.9 \times 10^4 \text{ s}$}, we expect approximately \mbox{$5.1 \times 10^{-7}$} and \mbox{$3.8 \times 10^{-6}$} counts for triple high-energy $\alpha$ summing and quadruple low-energy $\alpha$ summing, respectively. Thus, this $\alpha$ summing is not expected to have caused the five high-energy events.

Finally, the possibility of fission fragments being detected at low energy due to energy loss in the detector is considered. It must be noted that no fission fragments were measured in the \mbox{$A = 230$} spectrum. In order to approximate the probability of a fission fragment being measured in the 20 to 40 MeV range within a spectrum of similar proton number, a $^{202}$Fr spectrum from a 44-hour measurement at ISOLDE (CERN) from another study of $\beta$-delayed fission of proton-rich nuclei~\cite{ghysphd} was used. This experiment measured a sufficient amount of fission fragments in order to perform a fit to the probability density function of the fission fragment statistics. Integrating the probability density function over the 20 to 40 MeV range yields a probability of $4.4 \times 10^{-11}$ for a fission fragment being measured in this energy range. This estimate should be considered as an upper limit, as it was based on a spectrum with sufficient fission-fragment statistics to perform an analysis, while the \mbox{$A = 230$} spectrum did not feature any fission fragments. Therefore, the probability of the five high-energy events being fission fragments detected at a lower energy is negligible.

\subsection{Nuclei of a possible cluster emission}
\label{identifyingnuclei}
Continuing under the assumption that the five high-energy events can be attributed to cluster emission, we can determine the cluster nuclei most likely to be responsible for the events, as well as the parent isotope from which they would have been emitted. 

Each of the five high-energy events were measured several hours into the decay part of the measurement, therefore, $^{230}$Fr with its half-life of \mbox{$T_{1/2}$ = 19.1 s}~\cite{MORSE2024259} could be excluded as a parent isotope, because it would have decayed to a negligible level soon after the implantation ended.

$^{230}$Ra or $^{230}$Ac cannot be excluded as parent isotopes in this way as the isotopes are in secular equilibrium~\cite{PhysRevC.111.065803}. Nonetheless, the nature of cluster radioactivity as a decay mode heavily determined by shell and pairing effects of the residual nuclei influences which decays are likely to be observed. While the absence of these effects does not mean a decay channel is impossible, it does mean the half-lives of such a decay are likely to be less competitive compared to the favored cluster decay channels.

$^{230}$Ac is an odd-odd nucleus, meaning the cluster and the residual nucleus cannot both be even-even nuclei. Thus they miss out on pairing effects that both products of the even-even $^{230}$Ra would benefit from. The hindrance of cluster emission introduced by an odd-odd emitter is also reflected in the fact that among the available cluster emission measurements, the known emitters with an odd proton number $Z$ have an even neutron number $N$~\cite{bonetti_guglielmetti_2007}. No cluster emission has been detected from odd-odd nuclei. For these reasons, the measurement of cluster emission from $^{230}$Ra is expected to be more likely than from $^{230}$Ac.

Based on decay Q-value, shell effects (through the proximity of the residuals to magic nuclei, especially to the doubly magic $^{208}$Pb), and nuclear pairing effects of the residuals, the most likely cluster-daughter pair for cluster emission from $^{230}$Ra was identified. Most advantageous is the emission of $^{22}$O, resulting in a daughter nucleus of $^{208}$Hg. The former exhibits some magicity due to a sub-shell closure, as well as the magic proton number $Z$ = 8. The daughter nucleus needs two $\beta^-$ decays to reach the doubly magic $^{208}$Pb. The Q-value of this decay is 38.5(1) MeV, which lies within the 20 to 40 MeV range of the five high-energy events.

Another possibility is the emission of $^{24}\mathrm{O}$, leading to a residual of $^{206}$Hg, with a lower decay Q-value at 36.7(2) MeV, remaining within the expected range. Moreover $^{24}\mathrm{O}$ can be considered doubly magic~\cite{HOFFMAN200917} and $^{206}\mathrm{Hg}$ has a magic neutron number. However, the decay chain for $^{24}$O to reach stability consists of four $\beta^-$ decays, and despite its doubly magic status, it has not been experimentally observed as an emitted cluster so far~\cite{bonetti_guglielmetti_2007, sridhar_atlas_2020}.

\begin{table}[]
    \centering
    \caption{Estimated branching ratio (BR) and half-life for cluster emission from $^{230}\mathrm{Ra}$, under the assumption that the high-energy events are cluster emission measurements (first two rows). BR and $T_{1/2}$ of cluster emission from $^{230}$Ra calculated with each of the four analytical formulas RRF~\cite{SAXENA2024122867}, MBKAG~\cite{JAIN2023122597}, UDL~\cite{PhysRevLett.103.072501}, and NGN~\cite{Qi_2023} (final four rows). See text (and in particular Sect.~\ref{clusteremissionhalflives}) for more details.}
    \begin{tabular}{c|c|c}
    \hline 
         & BR & $\mathrm{T}_{1/2}$ (s)\\
         \hline
      5 events & ($4.25 \pm 1.94)\times \mathrm{10}^{-9}$ & ($1.31 \pm 0.60)\times \mathrm{10}^{12}$\\
      \hline
      RRF & $10^{-23}$ & $10^{27}$\\
      \hline
      MBKAG & $10^{-26}$ & $10^{30} $\\
      \hline
      UDL & $10^{-29}$ & $10^{33}$\\
      \hline
      NGN & $10^{-28}$ & $10^{32}$\\
      \hline
    \end{tabular}
    \label{tab:halflives}
\end{table}

Using the same criteria, we identified $^{22}\mathrm{O}$ as the most advantageous cluster to be emitted from $^{230}\mathrm{Ac}$, giving rise to the residual nucleus of $^{208}\mathrm{Tl}$. However, this decay is expected to be hindered, thus making it unlikely for $^{230}$Ac to be the parent isotope. For even-even emitters, like $^{230}$Ra, cluster emission proceeds via a $0^+ \rightarrow 0^+$ transition. On the other hand, for emitters with an odd mass number, the effects due to spin-change mean the decay will often go into an excited state of the daughter~\cite{bonetti_guglielmetti_2007}. While decays into excited states of the residual nuclei were not specifically analyzed in the present work, we note that for the emission of $^{22}\mathrm{O}$ from $^{230}\mathrm{Ac}$, assuming the ground state spin of $\mathrm{1}^+$~\cite{MORSE2024259}, neither the ground state, nor the excited states of $^{208}\mathrm{Tl}$ exhibit good overlap, thus severely hindering such a transition.

Finally, we calculated the branching ratio and half-life for cluster emission assuming either the five high-energy events are cluster emission from $^{230}\mathrm{Ra}$, regardless of which cluster is being emitted. The results of the calculation can be found in Table \ref{tab:halflives} and are discussed in Sect. \ref{clusteremissionhalflives}. 

\section{Cluster emission models for r-process calculation}
\label{rprocesstheory}
The long partial half-life ($\sim 10^{12} \text{ s}$) found for potential cluster emission from $^{230}\mathrm{Ra}$ in Sect. \ref{clusteremission} exemplifies the fact that cluster emission has so far been observed as an exotic decay mode limited to the actinide region. No cluster emission has been detected thus far for heavier elements, but a theoretical investigation in the super-heavy region~\cite{SAXENA2024122867} reported half-lives of cluster emission comparable to $\alpha$-particle emission. In this section we describe the calculations for the rates of cluster emission for heavy and superheavy nuclei, and include these in the calculation of r-process nucleosynthesis in the ejecta of a NS merger.

\subsection{Half-life models and calculation framework}

In order to calculate the rates of cluster emission for nuclei relevant to the r-process, four existing analytical formulas relating decay Q-value to cluster emission half-life were selected. These were the Universal Decay Law (UDL)~\cite{PhysRevLett.103.072501}, the Modified BKAG formula (MBKAG, where the initials refer to the authors' names)~\cite{JAIN2023122597}, the New Geiger-Nuttal law (NGN)~\cite{Qi_2023}, and the Refitted Royer Formula (RRF)~\cite{SAXENA2024122867}. Each of these formulas is based on describing cluster emission as an $\alpha$-decay-like process, considering the cluster as a preformed particle tunneling through a nuclear and Coulomb potential barrier. Such formulas are generally derived from cluster emission-specific modifications to the Geiger-Nuttall law. 

The UDL is an extension of the relation between half-life and Q-value for $\alpha$ decay assumed to also apply to cluster emission. It is used here for its ability to reproduce half-lives in the superheavy region predicted by models such as the Generalized Liquid Drop Model (GLDM) which includes fission-like mechanisms. Therefore, it may be appropriate to describe the heavy and superheavy region of the nuclear chart~\cite{SAXENA2024122867}. The MBKAG formula is based on the formula presented by Balasubramaniam et al.~\cite{PhysRevC.70.017301}, modified by Jain et al.~\cite{JAIN2023122597} to include angular momentum and isospin-dependent terms. In a comparison of several analytical formulas of cluster decay half-lives, MBKAG produced the lowest RMS and reduced $\chi^2$ compared to all other modified formulas~\cite{JAIN2023122597}. The NGN law is itself derived from the BKAG formula, aiming to be a Geiger-Nuttall law for describing cluster radioactivity~\cite{Qi_2023}. Finally, the RRF was presented by Saxena et al.~\cite{SAXENA2024122867} with the express purpose of providing theoretical cluster emission half-lives for superheavy nuclei, based on the Royer formula \cite{ROYER2001182}. In order to bridge the gap between the available experimental data in the actinide region and the superheavy region targeted by the RRF, they include 54 cluster emission half-lives of superheavy nuclei (104$\leq Z\leq$118) predicted using the GLDM in refitting the Royer formula~\cite{SAXENA2024122867}. Saxena et al. argue that both the UDL and RRF have predictive power in the superheavy region due to the inclusion of fission mechanisms in their formulas. MBKAG and NGN then serve as a comparison with the conventional formulas that are fitted exclusively to the actinide region.

Each formula takes the Q-value of the decay as input. The atomic masses used here in the calculation of these Q-values were taken from experimental data where available~\cite{Wang_2021}, and from the BSkG3 nuclear energy density functional model otherwise~\cite{grams_skyrme-hartree-fock-bogoliubov_2023}.

Within this paper, cluster emission was studied in terms of its effects on r-process nucleosynthesis in the astrophysical site of a NS merger. This influenced the selection of nuclei for which the cluster emission half-lives were calculated, namely for all elements with 82\textless $Z \leq$118. Primarily neutron-rich isotopes of these elements are predicted to be relevant to the r-process and will thus be included in the network. The full range of the group of nuclei relevant to the r-process can be seen in Figs.~\ref{fig:RRFUDLhalflives} and  \ref{fig:RRFUDLhalflivesorder}, where all isotopes on the neutron-rich side of the r-process network limit were included in the nucleosynthesis calculations.
For each nuclide, cluster emission was calculated for a decay leading to a residual of $^{208}$Pb or its neighboring nuclei ($Z \pm 2$ and/or $N \pm 2$), defining $Z_D \geq 80$ and $Z_C = Z - Z_D$, where subscript $C$ refers to the emitted cluster and $D$ to the residual daughter nucleus.

Since it is not known beforehand which isotope of $Z_C$ will be preferentially emitted from the parent isotope, the cluster emission half-lives were calculated for a range of isotopes of particular $Z_C$ that correspond to $A_S - 2 < A_C < A_S + 12$, where $A_S$ refers to the mass number of the lightest stable isotope of a cluster with a particular $Z_C$. For the heaviest parent nuclei ($Z \geq 100$), the range was expanded to $A_C < A_S + 23$. These ranges were chosen in order to include and extend the cluster isotopes whose emission has been observed experimentally, as those tend to be either a stable or a neutron-rich isotope. 

It must be noted that since these analytical formulas use an $\alpha$-decay-like description of cluster emission, the prediction is not expected to be accurate for odd mass number nuclei. Their decays can be strongly hindered depending on the nuclear structure of the residual nucleus~\cite{BONETTI199332}. Therefore, while the cluster emission half-lives will be calculated for a nuclear network of both even- and odd-$A$ nuclei, it is unclear whether the results for each odd-$A$ nucleus are reliable, as individual shell and pairing effects can have a severe impact on which emissions of a cluster will occur in reality. Ideally, when applied to exotic neutron-rich nuclei, the calculation of cluster emission half-lives should be based as much as possible on sound microscopic predictions, such as those obtained within self-consistent Gogny Hartree-Fock-Bogolyubov calculations~\cite{b7q7-925c}. Such calculations are however not available for the thousands of neutron-rich nuclei included in r-process reaction networks. For this reason, the present exploratory study on the impact of cluster emission on the r-process nucleosynthesis considers the few simple analytical formulas available nowadays.

\subsection{The r-process nucleosynthesis}
In order to study the impact of cluster emission on the \mbox{r-process} nucleosynthesis, we consider individual trajectories extracted from the NS merger simulation sym-n1-a6 presented in Ref.~\cite{Just_2023} corresponding to a 1.375-1.375 $M_\odot$ binary system. We only consider trajectories with a very low electron fraction ($Y_e$), thus indicating a neutron-rich composition, to favour the production of very heavy species that are potential candidates for cluster emission. In Sect.~\ref{results}, we will illustrate the case of a trajectory with $Y_e = 0.02$ only. The full extent of the r-process nucleosynthesis post-processing step is described in Refs.~\cite{Goriely_2011},~\cite{goriely_fundamental_2015}, and~\cite{10.1093/mnras/stab3393}. 

The nuclear physics input, both in the \mbox{r-process} nucleosynthesis calculations and in the comparison with the cluster emission rates calculated in this work, includes partial half-lives of spontaneous fission, taken from Ref.~\cite{SANCHEZFERNANDEZ2026140287} when not known experimentally. The \mbox{$\beta$-decay} processes which are not available experimentally are taken from the mean field plus relativistic QRPA calculation as described in Ref.~\cite{PhysRevC.93.025805}. The \mbox{$\alpha$-decay} rates from Ref.~\cite{Koura01082002} are taken into account. Cluster emission half-lives below the limit of $T_{1/2}(CE) \leq10 ^{17} \text{ s}$, the approximate age of the universe, are included in the reaction network, considering all open cluster emission decay channels. 

\section{Results \& Discussion}
\label{results}

\subsection{Comparison to $A = 230$ spectrum experimental data}

Before considering the cluster emission rates of heavy and superheavy nuclei on a broad scale, the branching ratios and half-lives for the emission of different clusters from $^{230}$Ra can be calculated with each of the four analytical formulas and compared to the experimentally obtained values from Sect.~\ref{230ra}. Table \ref{tab:halflives} lists the orders of magnitude for the emission of a cluster from $^{230}$Ra, for each of the four analytical formulas. When considering the emission of a specific cluster nucleus, the analytical formulas predict the shortest half-life for cluster decay from $^{230}$Ra to correspond to the emission of $^{14}$C, and the second shortest half-life for the emission of $^{22}$O, with the former being one, two, two, and four orders of magnitude shorter compared to the latter for RRF, MBKAG, UDL and NGN, respectively. While $^{14}$C is an isotope commonly emitted during cluster decay, for the case of $^{230}$Ra, the resulting daughter of $^{216}$Pb is located far from the valley of stability and is thus unlikely to be favored for cluster decay. The Q-value of such a decay would be 24.1(2) MeV, which is on the lower end of the 20-40 MeV range of measured energies for the five high-energy events. Therefore, the three highest-energy candidate events could not undergo this decay.

Overall, the analytical formulas thus agree qualitatively on $^{22}$O being the most likely isotope of O to be emitted from $^{230}$Ra. On the other hand, when considering the half-life of cluster emission from $^{230}$Ra as listed in Table~\ref{tab:halflives},
there is a very large quantitative difference between the order of magnitude of the predicted and experimental half-lives, with the shortest predicted value still being 15 orders of magnitude larger than the experimental one. Between the four formulas, the half-life predictions also vary by 6 orders of magnitude. This discrepancy suggests that the possibility of an alternative origin for the five high-energy events cannot be excluded. Another reason could be that factors having a significant effect on the cluster emission half-lives, such as nucleus deformation, are not fully included, or shell effects, which are included only indirectly through the Q-value~\cite{JAIN2023122597, SAXENA2024122867, Qi_2023, PhysRevLett.103.072501}. A study of the decay of $^{230}$Ra from a fully microscopic perspective could provide more clarity~\cite{b7q7-925c}.

\subsection{Cluster emission half-lives of the nuclear network}
\label{clusteremissionhalflives}

\begin{figure*}
    \centering
    \includegraphics[width=\linewidth]{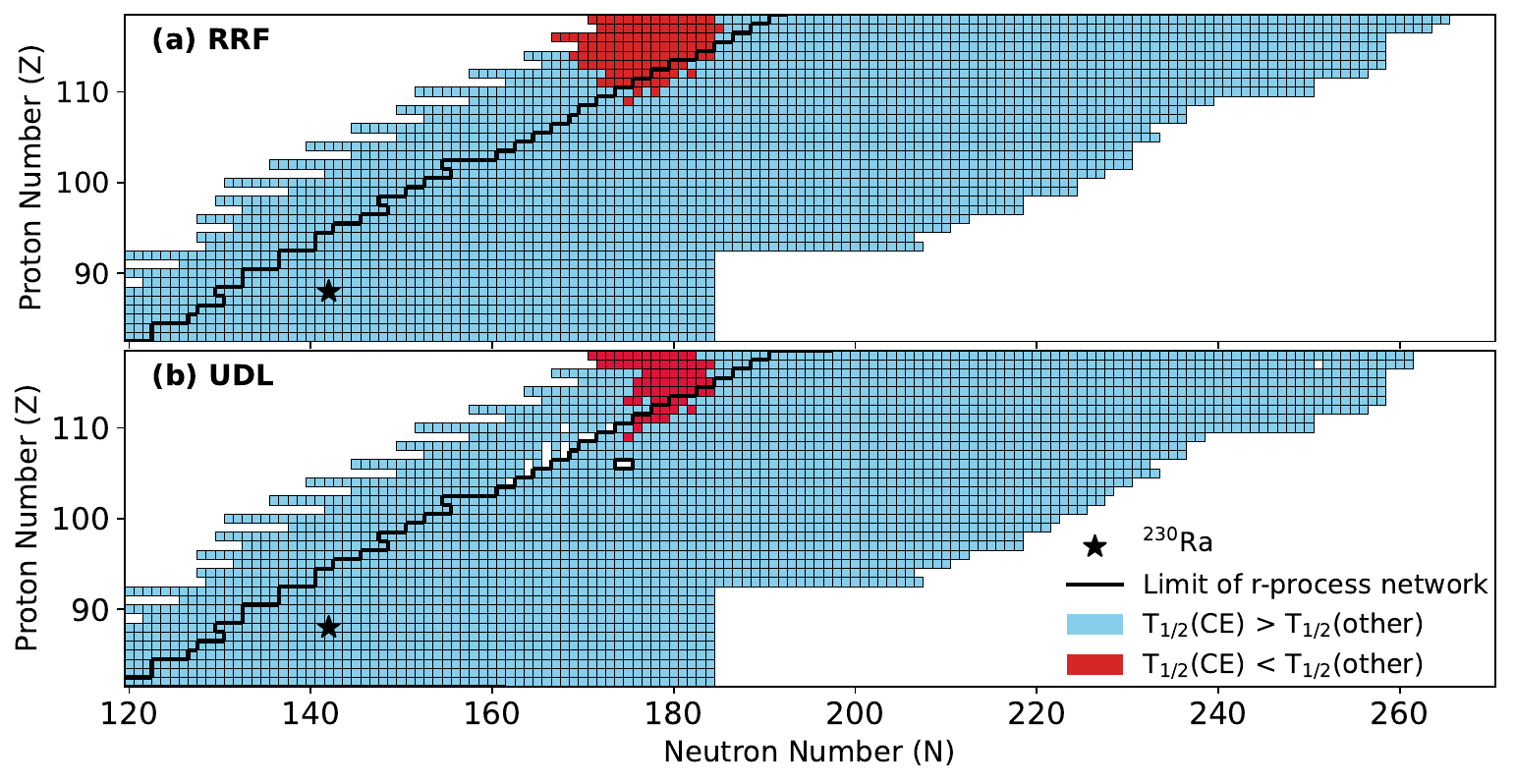}
    \caption{Comparison of the half-life of the cluster emission decay mode to the total half-life of the other decay modes, calculated using (a) RRF and (b) UDL. Nuclei on the neutron-rich side of the r-process network limit (black line) were considered in the r-process calculations. The studied isotope, $^{230}$Ra, is marked with a star.}
    \label{fig:RRFUDLhalflives}
\end{figure*}

\begin{figure*}
    \centering
    \includegraphics[width=\linewidth]{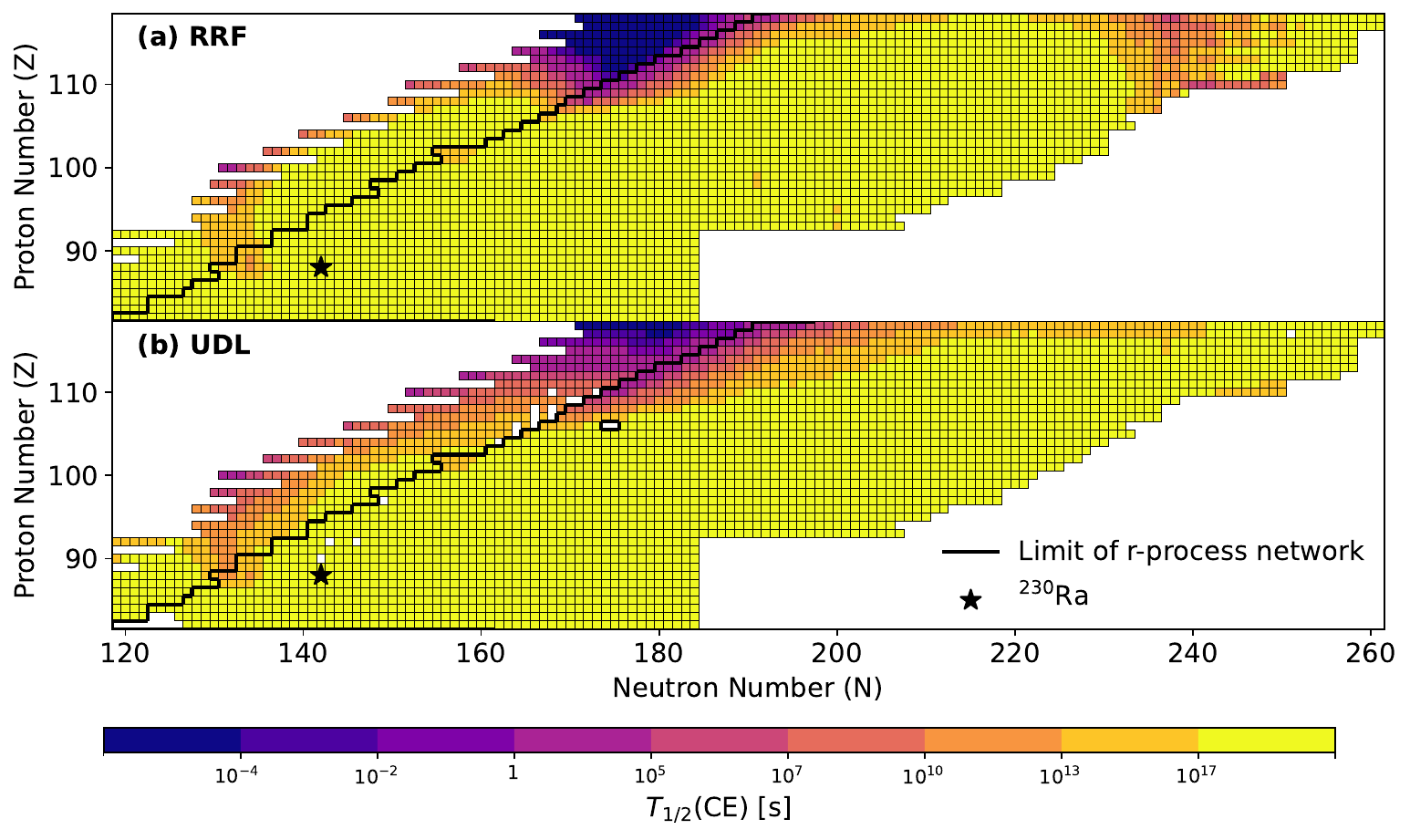}
    \caption{Total half-lives of cluster emission calculated using (a) RRF and (b) UDL. Nuclei on the neutron-rich side of the r-process network limit (black line) were considered in the r-process calculations. The studied isotope, $^{230}$Ra, is marked with a star.}
    \label{fig:RRFUDLhalflivesorder}
\end{figure*}

The partial half-life of cluster emission compared to the half-life of the sum of the other decay modes for all parent nuclei considered can be seen in Figs.~\ref{fig:RRFUDLhalflives}a and \ref{fig:RRFUDLhalflives}b for RRF and UDL models, respectively. In these and all subsequent figures, the nuclei included in the r-process network lie between the solid black line and the neutron drip line predicted by the BSkG3 mass model considered in the present calculations~\cite{grams_skyrme-hartree-fock-bogoliubov_2023}. Cluster emission half-lives are found to compete with other decay modes ($\beta$, $\alpha$, and spontaneous fission) for a small number of isotopes of elements above $Z$ = 110, in the neutron-deficient region of the r-process network. It can thus be noted that according to the predictions of both formulas, cluster emission is not expected to be the main decay mode of neutron-rich nuclei. Additionally, a smaller impact on the r-process by the UDL model can be anticipated, as evidenced by their longer half-lives (see Figs. \ref{fig:RRFUDLhalflivesorder}a-\ref{fig:RRFUDLhalflivesorder}b). A similar analysis was performed for the MBKAG and NGN formulas (see Sect. \ref{clusteremissionhalflives}). Both predict cluster emission not competing with the other decay modes (with the exception of $^{231,234}$Fm for MBKAG, two neutron-deficient isotopes that have not been measured yet).

The limited extent of nuclei for which cluster emission dominates the decay suggests also a limited impact on the r-process calculations. This can be analyzed in more detail by considering the order of magnitude of cluster emission half-lives as seen in Figs.~\ref{fig:RRFUDLhalflivesorder}a and \ref{fig:RRFUDLhalflivesorder}b. In both figures, one may note in the trans-actinide region, above $Z$ = 87 and around $N \approx$ 132, a region with low cluster emission half-lives relative to the surrounding nuclei. This can be recognized as the region from which the nuclei with experimentally measured half-lives of cluster emission originate.

Fission is predicted to become relevant to the r-process nucleosynthesis for nuclei with neutron and proton numbers above \mbox{$N \approx$ 160} and \mbox{$Z$ $\approx$ 95}, with major influence for \mbox{180 $\lesssim N \lesssim$ 200}~\cite{goriely_fundamental_2015}. Our calculations shown in Figs.~\ref{fig:RRFUDLhalflivesorder}a and \ref{fig:RRFUDLhalflivesorder}b do not indicate significant impact from cluster emission in this region. For the superheavy nuclei ($Z$ \textgreater 110), both the RRF and UDL models predict significantly shorter half-lives than in the rest of the chart. Saxena et al.~\cite{SAXENA2024122867} had previously reported half-lives below $10^{-2} \text{ s}$ for certain isotopes of Og with $284 \leq A \leq 306$. We find this conclusion extends to the region directly around these isotopes, for which the formulas predict half-lives below $10^{-4} \text{ s}$. The RRF indicates this region extends down to $Z =  112$. For UDL, the results show half-lives of 10$^{-4} \text{ s}$ only for isotopes of Ts and Og, while nuclei with $Z \geq 115$ have predicted half-lives below 1 s.

According to the predictions of the RRF, the half-lives increase rapidly when moving towards the neutron-rich side of the nuclear chart, as around $N \approx 200$ all half-lives are predicted to be longer than $10^{13} \text{ s}$. The increase predicted by the UDL is more gradual, with some isotopes of Og and Ts having half-lives between $10^5 - 10^7 \text{ s}$.

The only nuclei included in the r-process network with predicted half-lives below 1~s lie in the most neutron-deficient region. Only for nuclei with half-lives below this threshold is the inclusion of cluster emission expected to affect the r-process abundances. While these nuclei are included in the network, it is not expected they will be reached during the neutron irradiation. Indeed, the chain of $\beta^-$ decays needed to reach this region will likely be interrupted by spontaneous fission, which is the dominant decay mode for many nuclei on the path towards this region. Another feature that can be seen in Figs.~\ref{fig:RRFUDLhalflivesorder}a and \ref{fig:RRFUDLhalflivesorder}b is that there is a region of relatively low cluster emission half-lives centered around $Z = 115$ and $N = 240$. According to the RRF, the half-lives here are predicted to be as low as $10^5 \text{ s}$. For the UDL (see Fig.~\ref{fig:RRFUDLhalflivesorder}b), the feature is less pronounced, with some isotopes of Ds and Rg having a moderately lower half-life than those of the surrounding nuclei. The half-lives predicted in this region are comparable to some experimentally measured cluster emission half-lives, but significantly longer than what is expected to influence the r-process nucleosynthesis. It is not expected that the presence of this island of shorter half-lives will affect the isotopic abundance distribution.

\begin{figure*}
    \centering
    \includegraphics[width=\linewidth]{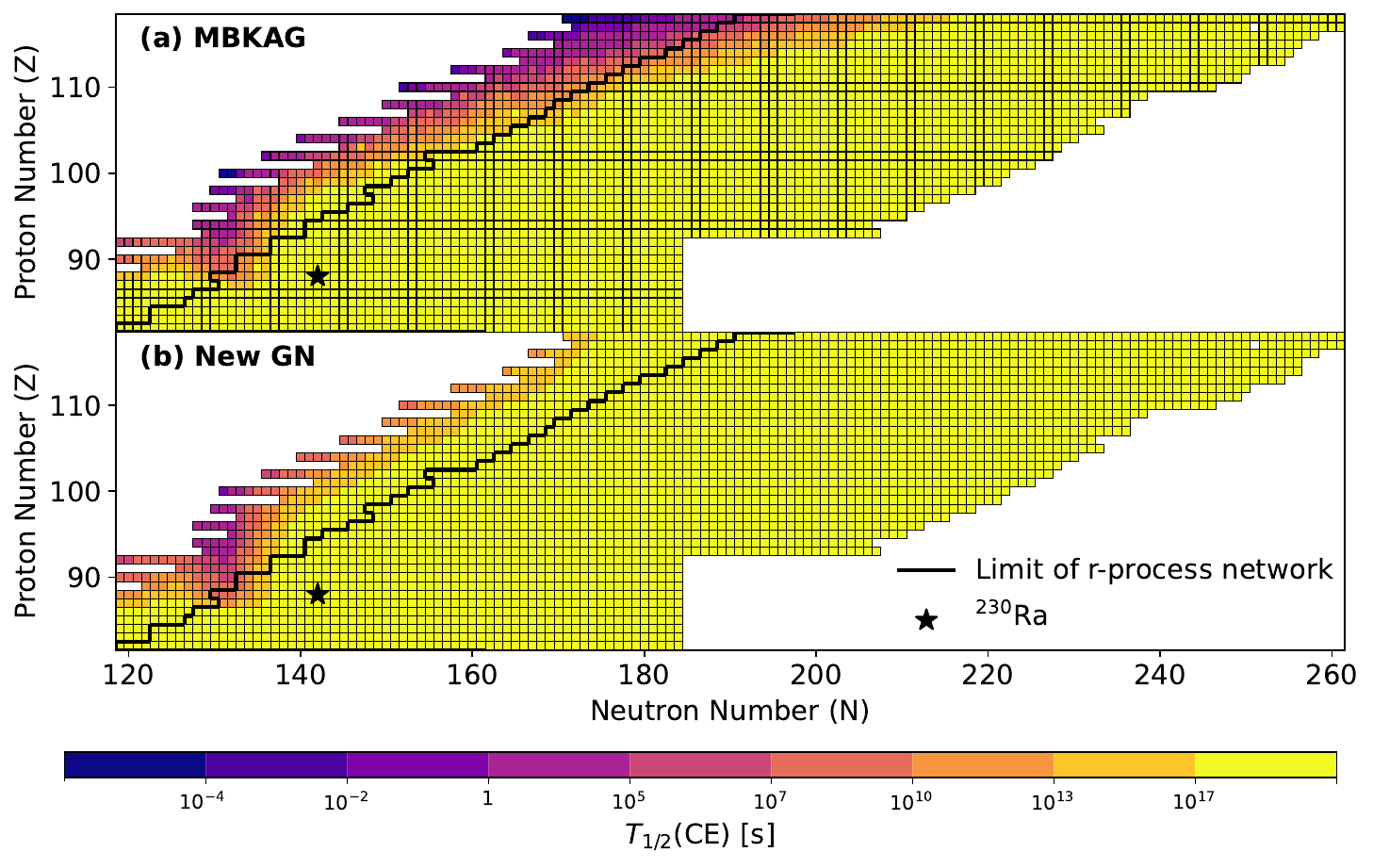}
    \caption{Total half-lives of cluster emission calculated using (a) MBKAG and (b) NGN. Nuclei on the neutron-rich side of the r-process network limit (black line) were considered in the r-process calculations. The studied isotope, $^{230}$Ra, is marked with a star.}
    \label{fig:MBKAGNGNhalflivesorder}
\end{figure*}

The half-life predictions for MBKAG and NGN are shown in Figs.~\ref{fig:MBKAGNGNhalflivesorder}a and \ref{fig:MBKAGNGNhalflivesorder}b. On the leftmost side of the chart, the same features can be identified as for UDL and RRF. This includes the region of experimentally measured cluster decays, to which the formulas have been fitted, and the shorter half-lives along the edge of the chart. In the neutron-deficient superheavy region, the half-lives predicted with MBKAG are moderately longer than those predicted with UDL and RRF. For the NGN formula, the difference is more pronounced, with no nuclei above $Z = 110$ having predicted cluster emission half-lives below $10^{10} \text{ s}$. Most notably, the island of lower cluster emission half-lives centered around $Z = 115$ and $N = 240$ does not appear for both MBKAG and NGN. Instead, these formulas predict a general trend of longer half-lives as one moves to the more neutron-deficient limit of the r-process network.

Some caveats have to be considered in assessing the reliability of the predictions from the MBKAG, NGN, UDL, and RRF models. First, MBKAG and NGN are fitted exclusively on the empirical cluster emission data in the actinide region, and they do not include any fission-like mechanisms, as opposed to the UDL and RRF. The divergence in results between the latter two and MBKAG and NGN is likely a result of this. A second general caveat is that since all four formulas require the Q-value of the decay as input, and the calculations were performed primarily for neutron-rich isotopes for which there are no experimental masses available, the predictions will also be influenced by the underlying nuclear mass model. The results presented in the current work were calculated using experimental mass excesses whenever available, supplemented by predictions with the BSkG3 model where not~\cite{grams_skyrme-hartree-fock-bogoliubov_2023}. The semi-microscopic character of this model is expected to provide greater predictive power in unexplored regions than a purely phenomenological model. Additionally, calculations were performed using D1M~\cite{PhysRevLett.102.242501}, FRDM12 \cite{MOLLER20161}, WS4 \cite{WANG2014215}, and HFB31 \cite{PhysRevC.93.034337} mass models. As the predictions using these models exhibit the same overall trends as BSkG3, only the latter results are presented here.

The calculation of half-lives for extremely neutron-rich isotopes will also be influenced by the underlying assumptions made in drafting each formula, a phenomenon that can be amplified when extrapolating results to very neutron-rich isotopes. For example, in Ref.~\cite{JAIN2023122597} the inclusion of the isospin term in MBKAG improved the fit to the available experimental data. However, regarding the neutron-rich side of the nuclear chart, there is no knowledge on whether the improvement holds in this different region. This term, which scales with the isospin, will be magnified by nearly a factor of two when comparing $^{378}$Og, a nucleus included here at the most neutron-rich edge of the nuclear network, to $^{294}$Og, the only isotope of Og which has been studied experimentally. If such a term fails to fully represent the underlying physical effect of isospin in the neutron-rich side of the nuclear chart, then deviations will be larger for this side of the nuclear chart.

Additionally, there is a large variance in predictions of cluster emission from the same parent nucleus by different formulas, which can be seen from Table \ref{tab:spread}. This table lists the predicted cluster emission half life for the aforementioned nucleus $^{294}$Og and for $^{356}$Ds, chosen as a representative example of extremely neutron-rich nuclei. While each of the four formulas is able to reproduce the available experimental cluster emission half-lives, the predictions become very disparate when extrapolating to the superheavy and neutron-rich region. The variance is especially notable for MBKAG and NGN, two models previously noted to provide a poor description of cluster emission in the superheavy region \cite{SAXENA2024122867}. This large variance illustrates the uncertainty associated with extrapolating phenomenological formulas beyond the experimentally known region and highlights the need for microscopic studies of cluster emission outside the trans-actinide region.

We note that RRF consistently predicts the shortest half-lives for cluster emission. One can therefore examine the predictions for the ``island of stability"~\cite{Oganessian_2012}. 
As an example, we take the case of $^{294}$Og. RRF predicts cluster emission to be the dominant decay mode, with a branching ratio of $\sim 1$ and a half-life of $10^{-16} \text{ s}$ for cluster emission. If this prediction were correct, cluster emission would overwhelmingly dominate the decay, strongly suppressing competing decay modes such as \mbox{$\alpha$-particle} emission and spontaneous fission. However, $\alpha$-particle emission has been measured experimentally, with a measured half-life of $0.58^{+0.44}_{-0.18}$ ms, and the upper limit for the branching ratio of spontaneous fission is reported to be 20\%~\cite{SINGH201970}. The prediction of a dominant cluster emission branch with a half-life of \mbox{$10^{-16}$} s is thus inconsistent with the observation of \mbox{$\alpha$-particle} emission. Therefore, we expect RRF to underestimate the cluster emission half-lives, at least in the region surrounding \mbox{$^{294}$Og}, and possibly in other regions of the nuclear chart as well. Consequently, based on long cluster emission half-lives for considered nuclei, it is unlikely that cluster emission will significantly affect the abundance distribution produced by the r-process.

\begin{table}[]
    \centering
    \caption{Predicted $T_{1/2}(CE)$ for $^{294}$Og and $^{356}$Ds, two nuclei included in the nuclear network, using the different formulas.}
    \begin{tabular}{c|c|c|c|c}
     & RRF [s] & UDL [s] & MBKAG [s] & NGN [s]\\
     \hline
     $^{294}$Og & $10^{-16}$ & $10^{-3}$ & $10^{19}$ & $10^{19}$\\
     \hline
     $^{356}$Ds & $10^9$ & $10^{14}$ & $10^{53}$ & $10^{84}$ \\
    \end{tabular}

    \label{tab:spread}
\end{table}

\subsection{Isotopic Abundance Distribution}

\begin{figure*}
    \centering
    \includegraphics[width=0.8\linewidth]{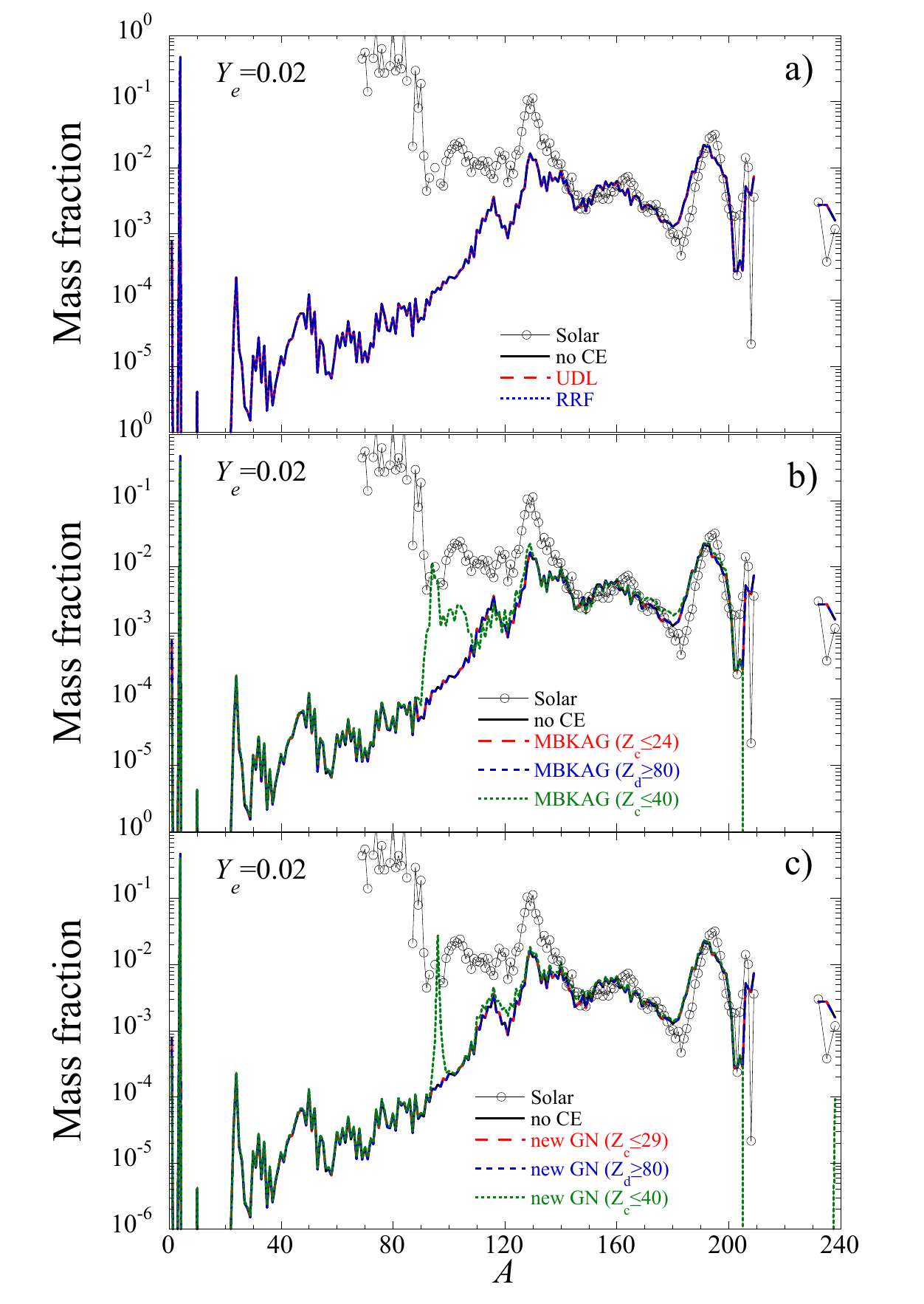}
    \caption{The r-process abundance distribution resulting from post-processing nucleosynthesis calculations on the NS merger model described in Sect. \ref{rprocesstheory}, after including the cluster emission rates calculated from (a) the UDL and RRF formula, (b) different ranges of $Z_C$ for MBKAG, and (c) different ranges of $Z_C$ for NGN. The solid black line, representing the calculation without cluster emission, is obscured by the red and blue curves in each panel.}
    \label{fig:abundances}
\end{figure*}

The results of the r-process post-processing calculations are shown in Fig. \ref{fig:abundances}. Figure \ref{fig:abundances}a confirms the earlier statement that the cluster emission half-lives are predicted to be too long to influence the r-process nucleosynthesis. The inclusion of both UDL and RRF cluster emission rates produces no change from the distribution calculated without cluster emission. Similarly, the case of cluster emission resulting in daughters with $Z_D \geq 80$ for MBKAG and NGN, indicated by a blue dashed line in Figs.~\ref{fig:abundances}b and \ref{fig:abundances}c produces no significant changes in the isotopic abundance distributions.

We also considered the case where cluster decay is calculated for emission of a cluster with $Z_C \leq 24$ and $Z_C \leq 40$ for MBKAG (Fig.~\ref{fig:abundances}b), as well as the cases for $Z_C \leq 29$ and $Z_C \leq 40$ for NGN (Fig.~\ref{fig:abundances}c). These calculations illustrate the limitations of MBKAG and NGN when extrapolated to the emission of heavier clusters. Heavy-cluster emission was included to investigate whether it could become relevant for the heaviest parent nuclei ($Z \gtrsim 110$). However, when the emission of clusters with $Z_C > 24$ (MBKAG) and $Z_C > 29$ (NGN) was allowed throughout the network, both formulas predicted a dominant decay mode similar to asymmetric fission for essentially all actinide and trans-actinide nuclei. For those nuclei whose decay properties have been measured experimentally, no such dominant heavy-cluster decay branch is observed, thus indicating these predictions are unphysical. For this reason, the calculations were limited to cluster emission resulting in a daughter with $Z_D \geq 80$. The case where the emission of heavy clusters was excluded entirely ($Z_C > 24$ and $Z_C > 29$ for MBKAG and NGN, respectively) is also presented for comparison.

MBKAG, with $Z_C \leq 24$, similar to $Z_D \geq 80$ predicts no effect on the isotopic abundance distribution, as seen in Fig.~\ref{fig:abundances}b. Indeed, even when the emission of heavier clusters ($Z_C > 24$) was calculated for the heaviest parent nuclei ($Z \geq 106$), the predicted half-lives were too long for cluster emission to be a significant decay mode. The distribution resulting from the calculation and inclusion of cluster decay with emission of a cluster with $Z_C \leq 40$ is also shown. As mentioned above, this distribution is not compatible with observations, and the resulting distribution predicts the destruction of all nuclei with $A > 205$ and a spike in production around $A = 100$. The same phenomenon is visible in Fig. \ref{fig:abundances}c for NGN, albeit with the upper limit on $Z_C \leq 29$ of the emitted cluster, at which the inclusion of heavier clusters no longer allows for the reproduction of the experimental cluster emission half-lives.

Ideally, extending the calculations to heavier emitted clusters should simply produce prohibitively long half-lives wherever such decays are not physically relevant. This behavior is obtained with UDL and RRF, but not with MBKAG and NGN. When imposing similar restrictions on the range of emitted clusters for each parent isotope, the predictions obtained for UDL and RRF did not change with respect to those of the standard case ($Z_D \geq 80$). For each case, regardless of whether the emission of a cluster with $Z_C > 24$ was permitted for nuclei with $Z > 82$, a spurious asymmetric fission channel as present for MBKAG and NGN was not predicted, and the experimental cluster emission half-lives could be reproduced.

\section{Conclusion}
\label{conclusion}
This work first reports five high-energy events in a spectrum of mass $A$ = 230 measured during an experimental campaign dedicated to $\beta$-delayed fission of $^{230,232,234}$Ac at ISOLDE (CERN). After exclusion of the possibilities of these events originating from contamination in the beam, electronic noise, $\alpha$ summing, or being very low energy fission fragments, the most likely parent-cluster pair is determined under the assumption that the high-energy events originate from cluster emission. The most likely parent-cluster combination was found to be $^{230}$Ra and $^{22}$O, resulting in a partial  half-life of $(1.31 \pm 0.60) \times 10^{12} \text{ s}$. Comparison with the half-lives predicted by the four analytical formulas indicates agreement on $^{22}$O being the most likely cluster to be emitted from $^{230}$Ra in the event of a cluster decay, but models predict a half-life for this decay longer by at least 15 orders of magnitude. Moreover, theoretical predictions by the used formulas spread over 6 orders of magnitude. Future work could include a dedicated study of cluster emission from the isotopes contained in the mass $A$ = 230 beam, with special attention to $^{230}$Ra, which could provide additional and well-calibrated data in order to determine the origin of the five high-energy events. Additionally, analyzing cluster emission within $^{230}$Ra from a fully microscopic theory such as the approach taken by Warda et al. in Ref.~\cite{warda2011} could help determine if cluster emission is at all likely from this nucleus.

Furthermore, the possible influence of cluster emission on the r-process is studied by calculating the rates of cluster emission from the analytical formulas UDL~\cite{PhysRevLett.103.072501}, MBKAG~\cite{JAIN2023122597}, NGN~\cite{Qi_2023}, and RRF~\cite{ROYER2001182} relating decay Q-value to partial half-life. It was found that besides the immediate trans-actinide region where cluster emission was studied experimentally, cluster emission half-lives are predicted to have a measurable influence only for the heaviest nuclei, centered around $^{294}$Og. Additionally, the predictions vary according to the four formulas. NGN predicts an absence of cluster emission for the nuclei relevant to the r-process. The predictions based on MBKAG indicate that only some nuclei at the neutron-deficient limit of the r-process reaction network have cluster emission half-lives shorter than the approximate age of the universe, but none with half-lives short enough to compete with the $\beta$, $\alpha$, and fission channels. UDL and RRF predict shorter cluster emission half-lives overall and a region of shorter half-lives for very neutron-rich nuclei with $Z >106$. Based on a comparison to the experimentally measured isotope $^{294}$Og, RRF is expected to underestimate the cluster emission half-lives, and thus overestimate the effect of cluster emission on the r-process. The discrepancy in predictions between MBKAG and NGN on the one hand, and UDL and RRF on the other, is in line with expectations, as MBKAG and NGN are not expected to produce accurate results for heavy nuclei~\cite{ SAXENA2024122867}. Due to RRF predicting the shortest half-lives among the four formulas, its predictions can be used as an upper limit for the effect of cluster emission on the r-process. We incorporated the cluster emission half-lives into the calculations of r-process nucleosynthesis in a NS merger model. However, even for the upper limit of RRF, cluster emission did not have an impact on the resulting abundance distribution. Such a conclusion remains, however, based on the few simple analytical formulas available to describe cluster emission probabilities. Such formulas are adjusted to reproduce experimental data of actinides and can be questioned when extrapolated far away in the neutron-rich region of astrophysical interest. Future microscopic mean-field calculations~\cite{warda2011, b7q7-925c} could provide a deeper insight on the possible cluster emission by neutron-rich nuclei.

\bmhead{Acknowledgements}
The authors would like to thank the ISOLDE Collaboration and technical teams at CERN for their support and the GSI Target Laboratory for the carbon foils used in the experimental campaign. The present research benefited from computational resources made available on Lucia, the Tier-1 supercomputer of the Walloon Region, infrastructure funded by the Walloon Region under the grant agreement n°1910247.

\section*{Declarations}

\textbf{Funding:} NT is a FRIA grantee of the Fonds de la Recherche Scientifique - FNRS funded by the French-speaking community of Belgium. SG is a Fonds de la Recherche Scientifique - FNRS research associate. This work has received funding from Fonds Wetenschappelijk Onderzoek (Belgium) and Fonds de la Recherche Scientifique - FNRS (Belgium) under the Excellence Of Science (EOS) program (Grant No. 30468642, No. 40007501, and No. O000422), from the Research Foundation Flanders under Projects I002619N, I002919N, and I012420N of the International Research Infrastructure, from a Fond Wetenschappelijk Onderzoek fellowship for fundamental research (Contract No. 1167324N), from the KU Leuven BOF (C14/22/104), from the European Commission’s Horizon Europe ERC Consolidator Grant 101088504 (NSHAPE), from the Slovak Research and Development Agency (Contract No. APVV-22-0282), and the Slovak grant agency VEGA (Contract No. 1/0019/25), and from the Romanian National Authority for Research through the Nucleu Project No. \mbox{PN 23 21 01 02} and the Institute of Atomic Physics through the CERN/ISOLDE grant.

\bibliography{sn-bibliography}

@article{ARNOULD2020103766,
title = {Astronuclear Physics: A tale of the atomic nuclei in the skies},
journal = {Prog. Part. Nucl. Phys.},
volume = {112},
pages = {103766},
year = {2020},
issn = {0146-6410},
doi = {https://doi.org/10.1016/j.ppnp.2020.103766},
url = {https://www.sciencedirect.com/science/article/pii/S0146641020300132},
author = {M. Arnould and S. Goriely}
}

@ARTICLE{1965ApJS...11..121S,
       author = {{Seeger}, Philip A. and {Fowler}, William A. and {Clayton}, Donald D.},
        title = "{Nucleosynthesis of Heavy Elements by Neutron Capture.}",
      journal = {Astrophys. J. Suppl.},
         year = 1965,
        month = feb,
       volume = {11},
        pages = {121},
          doi = {10.1086/190111},
       adsurl = {https://ui.adsabs.harvard.edu/abs/1965ApJS...11..121S}
}

@article{ARNOULD200797,
title = {The r-process of stellar nucleosynthesis: Astrophysics and nuclear physics achievements and mysteries},
journal = {Phys. Rep.},
volume = {450},
number = {4},
pages = {97-213},
year = {2007},
issn = {0370-1573},
doi = {https://doi.org/10.1016/j.physrep.2007.06.002},
url = {https://www.sciencedirect.com/science/article/pii/S0370157307002438},
author = {M. Arnould and S. Goriely and K. Takahashi}
}

@ARTICLE{1999A&A...342..881G,
       author = {{Goriely}, S.},
        title = "{Uncertainties in the solar system r-abundance distribution}",
      journal = {Astron. Astrophys.},
         year = 1999,
        month = feb,
       volume = {342},
        pages = {881-891},
       adsurl = {https://ui.adsabs.harvard.edu/abs/1999A&A...342..881G}
}

@ARTICLE{1974ApJ...192L.145L,
       author = {{Lattimer}, J.~M. and {Schramm}, D.~N.},
        title = "{Black-Hole-Neutron-Star Collisions}",
      journal = {Astrophys. J. Lett.},
         year = 1974,
        month = sep,
       volume = {192},
        pages = {L145},
          doi = {10.1086/181612},
       adsurl = {https://ui.adsabs.harvard.edu/abs/1974ApJ...192L.145L}
}

@ARTICLE{1976ApJ...210..549L,
       author = {{Lattimer}, J.~M. and {Schramm}, D.~N.},
        title = "{The tidal disruption of neutron stars by black holes in close binaries.}",
      journal = {Astrophys. J.},
         year = 1976,
        month = dec,
       volume = {210},
        pages = {549-567},
          doi = {10.1086/154860},
       adsurl = {https://ui.adsabs.harvard.edu/abs/1976ApJ...210..549L}
}

@ARTICLE{1982ApL....22..143S,
       author = {{Symbalisty}, E. and {Schramm}, D.~N.},
        title = "{Neutron Star Collisions and the r-Process}",
      journal = {Astrophys. Lett.},
         year = 1982,
        month = jan,
       volume = {22},
        pages = {143},
       adsurl = {https://ui.adsabs.harvard.edu/abs/1982ApL....22..143S}
}

@ARTICLE{2005NuPhA.758..587G,
       author = {{Goriely}, S. and {Demetriou}, P. and {Janka}, H. -Th. and {Pearson}, J.~M. and {Samyn}, M.},
        title = "{The r-process nucleosynthesis: a continued challenge for nuclear physics and astrophysics}",
      journal = {Nucl. Phys. A},
         year = 2005,
        month = jul,
       volume = {758},
        pages = {587-594},
          doi = {10.1016/j.nuclphysa.2005.05.107},
archivePrefix = {arXiv},
       eprint = {astro-ph/0410429},
 primaryClass = {astro-ph},
       adsurl = {https://ui.adsabs.harvard.edu/abs/2005NuPhA.758..587G}
}

@article{PhysRevLett.102.242501,
  title = {First {G}ogny-{H}artree-{F}ock-{B}ogoliubov Nuclear Mass Model},
  author = {Goriely, S. and Hilaire, S. and Girod, M. and P\'eru, S.},
  journal = {Phys. Rev. Lett.},
  volume = {102},
  issue = {24},
  pages = {242501},
  numpages = {4},
  year = {2009},
  month = {Jun},
  publisher = {American Physical Society},
  doi = {10.1103/PhysRevLett.102.242501},
  url = {https://link.aps.org/doi/10.1103/PhysRevLett.102.242501}
}

@article{WANG2014215,
title = {Surface diffuseness correction in global mass formula},
journal = {Physics Letters B},
volume = {734},
pages = {215-219},
year = {2014},
issn = {0370-2693},
doi = {https://doi.org/10.1016/j.physletb.2014.05.049},
url = {https://www.sciencedirect.com/science/article/pii/S037026931400358X},
author = {Ning Wang and Min Liu and Xizhen Wu and Jie Meng}
}

@article{KONDEV2021509,
title = {Nuclear Data Sheets for {A}=203},
journal = {Nuclear Data Sheets},
volume = {177},
pages = {509-699},
year = {2021},
issn = {0090-3752},
doi = {https://doi.org/10.1016/j.nds.2021.09.002},
url = {https://www.sciencedirect.com/science/article/pii/S0090375221000533},
author = {F.G. Kondev}
}

@article{KONDEV2025346,
title = {Recommended Nuclear Structure and Decay Data for {A}=206 Isobars},
journal = {Nuclear Data Sheets},
volume = {201},
pages = {346-607},
year = {2025},
issn = {0090-3752},
doi = {https://doi.org/10.1016/j.nds.2025.03.002},
url = {https://www.sciencedirect.com/science/article/pii/S0090375225000249},
author = {F.G. Kondev}
}

@article{PhysRevC.93.034337,
  title = {Further explorations of {S}kyrme-{H}artree-{F}ock-{B}ogoliubov mass formulas. {XVI}. Inclusion of self-energy effects in pairing},
  author = {Goriely, S. and Chamel, N. and Pearson, J. M.},
  journal = {Phys. Rev. C},
  volume = {93},
  issue = {3},
  pages = {034337},
  numpages = {11},
  year = {2016},
  month = {Mar},
  publisher = {American Physical Society},
  doi = {10.1103/PhysRevC.93.034337},
  url = {https://link.aps.org/doi/10.1103/PhysRevC.93.034337}
}

@article{MOLLER20161,
title = {Nuclear ground-state masses and deformations: {FRDM}(2012)},
journal = {Atomic Data and Nuclear Data Tables},
volume = {109-110},
pages = {1-204},
year = {2016},
issn = {0092-640X},
doi = {https://doi.org/10.1016/j.adt.2015.10.002},
url = {https://www.sciencedirect.com/science/article/pii/S0092640X1600005X},
author = {P. Möller and A.J. Sierk and T. Ichikawa and H. Sagawa}
}

@article{RevModPhys.93.015002,
  title = {Origin of the heaviest elements: The rapid neutron-capture process},
  author = {Cowan, John J. and Sneden, Christopher and Lawler, James E. and Aprahamian, Ani and Wiescher, Michael and Langanke, Karlheinz and Mart\'{\i}nez-Pinedo, Gabriel and Thielemann, Friedrich-Karl},
  journal = {Rev. Mod. Phys.},
  volume = {93},
  issue = {1},
  pages = {015002},
  numpages = {85},
  year = {2021},
  month = {Feb},
  publisher = {American Physical Society},
  doi = {10.1103/RevModPhys.93.015002},
  url = {https://link.aps.org/doi/10.1103/RevModPhys.93.015002}
}

@article{KAJINO2019109,
title = {Current status of r-process nucleosynthesis},
journal = {Prog. Part. Nucl. Phys.},
volume = {107},
pages = {109-166},
year = {2019},
issn = {0146-6410},
doi = {https://doi.org/10.1016/j.ppnp.2019.02.008},
url = {https://www.sciencedirect.com/science/article/pii/S0146641019300201},
author = {T. Kajino and W. Aoki and A.B. Balantekin and R. Diehl and M.A. Famiano and G.J. Mathews}
}

@article{Woosley1994,
author = {Woosley, S. and Wilson, J. and Mathews, G. and Hoffman, R. and Meyer, B.},
year = {1994},
month = {10},
pages = {},
title = {The r-process and neutrino-heated supernova ejecta},
volume = {433},
journal = {Astrophys. J.},
doi = {10.1086/174638}
}

@article{Wanajo_2012,
doi = {10.1088/0004-637X/746/2/180},
url = {https://doi.org/10.1088/0004-637X/746/2/180},
year = {2012},
month = {feb},
publisher = {The American Astronomical Society},
volume = {746},
number = {2},
pages = {180},
author = {Wanajo, Shinya and Janka, Hans-Thomas},
title = {The r-PROCESS IN THE NEUTRINO-DRIVEN WIND FROM A BLACK-HOLE TORUS},
journal = {Astrophys. J.}
}

@article{WANAJO2006676,
title = {r-process calculations and Galactic chemical evolution},
journal = {Nucl. Phys. A},
volume = {777},
pages = {676-699},
year = {2006},
note = {Special Issue on Nuclear Astrophysics},
issn = {0375-9474},
doi = {https://doi.org/10.1016/j.nuclphysa.2005.10.012},
url = {https://www.sciencedirect.com/science/article/pii/S0375947405011814},
author = {Shinya Wanajo and Yuhri Ishimaru}
}

@ARTICLE{2011ApJ...726L..15W,
       author = {{Wanajo}, Shinya and {Janka}, Hans-Thomas and {M{\"u}ller}, Bernhard},
        title = "{Electron-capture Supernovae as The Origin of Elements Beyond Iron}",
      journal = {Astrophys. J. Lett.},
         year = 2011,
        month = jan,
       volume = {726},
       number = {2},
          eid = {L15},
        pages = {L15},
          doi = {10.1088/2041-8205/726/2/L15},
archivePrefix = {arXiv},
       eprint = {1009.1000},
 primaryClass = {astro-ph.SR},
       adsurl = {https://ui.adsabs.harvard.edu/abs/2011ApJ...726L..15W}
}

@article{MARTIN20071583,
title = {Nuclear Data Sheets for {A} = 208},
journal = {Nuclear Data Sheets},
volume = {108},
number = {8},
pages = {1583-1806},
year = {2007},
issn = {0090-3752},
doi = {https://doi.org/10.1016/j.nds.2007.07.001},
url = {https://www.sciencedirect.com/science/article/pii/S0090375207000658},
author = {M.J. Martin}
}

@article{MORSE2026409,
title = {Nuclear Structure and Decay Data for {A}=216 Isobars},
journal = {Nuclear Data Sheets},
volume = {209},
pages = {409-498},
year = {2026},
issn = {0090-3752},
doi = {https://doi.org/10.1016/j.nds.2026.02.002},
url = {https://www.sciencedirect.com/science/article/pii/S0090375226000165},
author = {C. Morse}
}

@article{KONDEV2018382,
title = {Nuclear Data Sheets for {A}=217},
journal = {Nuclear Data Sheets},
volume = {147},
pages = {382-458},
year = {2018},
issn = {0090-3752},
doi = {https://doi.org/10.1016/j.nds.2018.01.002},
url = {https://www.sciencedirect.com/science/article/pii/S0090375218300024},
author = {F.G. Kondev and E.A. McCutchan and B. Singh and K. Banerjee and S. Bhattacharya and A. Chakraborty and S. Garg and N. Jovancevic and S. Kumar and S.K. Rathi and T. Roy and Jounghwa Lee and R. Shearman}
}

@article{CHEN2015373,
title = {Nuclear Data Sheets for {A} = 209},
journal = {Nuclear Data Sheets},
volume = {126},
pages = {373-546},
year = {2015},
issn = {0090-3752},
doi = {https://doi.org/10.1016/j.nds.2015.05.003},
url = {https://www.sciencedirect.com/science/article/pii/S0090375215000149},
author = {J. Chen and F.G. Kondev}
}

@article{BASUNIA2022475,
title = {Nuclear Data Sheets for {A}=213},
journal = {Nuclear Data Sheets},
volume = {181},
pages = {475-585},
year = {2022},
issn = {0090-3752},
doi = {https://doi.org/10.1016/j.nds.2022.03.002},
url = {https://www.sciencedirect.com/science/article/pii/S0090375222000096},
author = {M.S. Basunia}
}

@article{SHAMSUZZOHABASUNIA2014561,
title = {Nuclear Data Sheets for {A} = 210},
journal = {Nuclear Data Sheets},
volume = {121},
pages = {561-694},
year = {2014},
issn = {0090-3752},
doi = {https://doi.org/10.1016/j.nds.2014.09.004},
url = {https://www.sciencedirect.com/science/article/pii/S0090375214006589},
author = {M. {Shamsuzzoha Basunia}}
}

@article{SINGH2013661,
title = {Nuclear Data Sheets for {A} = 211},
journal = {Nuclear Data Sheets},
volume = {114},
number = {6},
pages = {661-749},
year = {2013},
issn = {0090-3752},
doi = {https://doi.org/10.1016/j.nds.2013.05.001},
url = {https://www.sciencedirect.com/science/article/pii/S0090375213000410},
author = {Balraj Singh and Daniel Abriola and Coral Baglin and Vivian Demetriou and Timothy Johnson and Elizabeth McCutchan and Gopal Mukherjee and Sukhjeet Singh and Alejandro Sonzogni and Jagdish Tuli}
}

@article{KONDEV2011707,
title = {Nuclear Data Sheets for {A} = 207},
journal = {Nuclear Data Sheets},
volume = {112},
number = {3},
pages = {707-853},
year = {2011},
issn = {0090-3752},
doi = {https://doi.org/10.1016/j.nds.2011.02.002},
url = {https://www.sciencedirect.com/science/article/pii/S0090375211000111},
author = {F.G. Kondev and S. Lalkovski}
}

@article{PhysRevLett.109.251104,
  title = {Charged-Current Weak Interaction Processes in Hot and Dense Matter and its Impact on the Spectra of Neutrinos Emitted from Protoneutron Star Cooling},
  author = {Mart\'{\i}nez-Pinedo, G. and Fischer, T. and Lohs, A. and Huther, L.},
  journal = {Phys. Rev. Lett.},
  volume = {109},
  issue = {25},
  pages = {251104},
  numpages = {5},
  year = {2012},
  month = {Dec},
  publisher = {American Physical Society},
  doi = {10.1103/PhysRevLett.109.251104},
  url = {https://link.aps.org/doi/10.1103/PhysRevLett.109.251104}
}

@article{Oganessian_2012,
doi = {10.1088/1742-6596/337/1/012005},
url = {https://doi.org/10.1088/1742-6596/337/1/012005},
year = {2012},
month = {feb},
publisher = {},
volume = {337},
number = {1},
pages = {012005},
author = {Oganessian, Yuri},
title = {Nuclei in the ``Island of Stability" of Superheavy Elements},
journal = {Journal of Physics: Conference Series}
}

@misc{lirias4273769,
author = {Bara, Silvia and Cocolios},
language = {eng},
title = {Experimental and theoretical studies of beta-delayed fission, PhD thesis},
address = {KU Leuven},
year = {2025},
}

@article{SANCHEZFERNANDEZ2026140287,
title = {Accurate spontaneous fission half-lives from a microscopic large-scale nuclear structure model},
journal = {Physics Letters B},
volume = {874},
pages = {140287},
year = {2026},
issn = {0370-2693},
doi = {https://doi.org/10.1016/j.physletb.2026.140287},
url = {https://www.sciencedirect.com/science/article/pii/S0370269326001413},
author = {A. Sánchez-Fernández and S. Bara and W. Ryssens and S. Goriely}
}

@article{PhysRevC.86.065803,
  title = {Medium modification of the charged-current neutrino opacity and its implications},
  author = {Roberts, L. F. and Reddy, Sanjay and Shen, Gang},
  journal = {Phys. Rev. C},
  volume = {86},
  issue = {6},
  pages = {065803},
  numpages = {10},
  year = {2012},
  month = {Dec},
  publisher = {American Physical Society},
  doi = {10.1103/PhysRevC.86.065803},
  url = {https://link.aps.org/doi/10.1103/PhysRevC.86.065803}
}

@article{Wang_2023,
doi = {10.3847/1538-4357/ace7b2},
url = {https://doi.org/10.3847/1538-4357/ace7b2},
year = {2023},
month = {aug},
publisher = {The American Astronomical Society},
volume = {954},
number = {2},
pages = {114},
author = {Wang, Tianshu and Burrows, Adam},
title = {Neutrino-driven Winds in Three-dimensional Core-collapse Supernova Simulations},
journal = {Astrophys. J.}
}

@ARTICLE{2017PhRvL.119p1101A,
       author = {{Abbott}, B.~P. and {Abbott}, R. and {Abbott}, T.~D. and {Acernese}, F. and {Ackley}, K. and {Adams}, C. and {Adams}, T. and {Addesso}, P. and {Adhikari}, R.~X. and {Adya}, V.~B. and {Affeldt}, C. and {Afrough}, M. and {Agarwal}, B. and {Agathos}, M. and {Agatsuma}, K. and {Aggarwal}, N. and {Aguiar}, O.~D. and {Aiello}, L. and {Ain}, A. and {Ajith}, P. and {Allen}, B. and {Allen}, G. and {Allocca}, A. and {Altin}, P.~A. and {Amato}, A. and {Ananyeva}, A. and {Anderson}, S.~B. and {Anderson}, W.~G. and {Angelova}, S.~V. and {Antier}, S. and {Appert}, S. and {Arai}, K. and {Araya}, M.~C. and {Areeda}, J.~S. and {Arnaud}, N. and {Arun}, K.~G. and {Ascenzi}, S. and {Ashton}, G. and {Ast}, M. and {Aston}, S.~M. and {Astone}, P. and {Atallah}, D.~V. and {Aufmuth}, P. and {Aulbert}, C. and {AultONeal}, K. and {Austin}, C. and {Avila-Alvarez}, A. and {Babak}, S. and {Bacon}, P. and {Bader}, M.~K.~M. and {Bae}, S. and {Bailes}, M. and {Baker}, P.~T. and {Baldaccini}, F. and {Ballardin}, G. and {Ballmer}, S.~W. and {Banagiri}, S. and {Barayoga}, J.~C. and {Barclay}, S.~E. and {Barish}, B.~C. and {Barker}, D. and {Barkett}, K. and {Barone}, F. and {Barr}, B. and {Barsotti}, L. and {Barsuglia}, M. and {Barta}, D. and {Barthelmy}, S.~D. and {Bartlett}, J. and {Bartos}, I. and {Bassiri}, R. and {Basti}, A. and {Batch}, J.~C. and {Bawaj}, M. and {Bayley}, J.~C. and {Bazzan}, M. and {B{\'e}csy}, B. and {Beer}, C. and {Bejger}, M. and {Belahcene}, I. and {Bell}, A.~S. and {Berger}, B.~K. and {Bergmann}, G. and {Bernuzzi}, S. and {Bero}, J.~J. and {Berry}, C.~P.~L. and {Bersanetti}, D. and {Bertolini}, A. and {Betzwieser}, J. and {Bhagwat}, S. and {Bhandare}, R. and {Bilenko}, I.~A. and {Billingsley}, G. and {Billman}, C.~R. and {Birch}, J. and {Birney}, R. and {Birnholtz}, O. and {Biscans}, S. and {Biscoveanu}, S. and {Bisht}, A. and {Bitossi}, M. and {Biwer}, C. and {Bizouard}, M.~A. and {Blackburn}, J.~K. and {Blackman}, J. and {Blair}, C.~D. and {Blair}, D.~G. and {Blair}, R.~M. and {Bloemen}, S. and {Bock}, O. and {Bode}, N. and {Boer}, M. and {Bogaert}, G. and {Bohe}, A. and {Bondu}, F. and {Bonilla}, E. and {Bonnand}, R. and {Boom}, B.~A. and {Bork}, R. and {Boschi}, V. and {Bose}, S. and {Bossie}, K. and {Bouffanais}, Y. and {Bozzi}, A. and {Bradaschia}, C. and {Brady}, P.~R. and {Branchesi}, M. and {Brau}, J.~E. and {Briant}, T. and {Brillet}, A. and {Brinkmann}, M. and {Brisson}, V. and {Brockill}, P. and {Broida}, J.~E. and {Brooks}, A.~F. and {Brown}, D.~A. and {Brown}, D.~D. and {Brunett}, S. and {Buchanan}, C.~C. and {Buikema}, A. and {Bulik}, T. and {Bulten}, H.~J. and {Buonanno}, A. and {Buskulic}, D. and {Buy}, C. and {Byer}, R.~L. and {Cabero}, M. and {Cadonati}, L. and {Cagnoli}, G. and {Cahillane}, C. and {Calder{\'o}n Bustillo}, J. and {Callister}, T.~A. and {Calloni}, E. and {Camp}, J.~B. and {Canepa}, M. and {Canizares}, P. and {Cannon}, K.~C. and {Cao}, H. and {Cao}, J. and {Capano}, C.~D. and {Capocasa}, E. and {Carbognani}, F. and {Caride}, S. and {Carney}, M.~F. and {Carullo}, G. and {Casanueva Diaz}, J. and {Casentini}, C. and {Caudill}, S. and {Cavagli{\`a}}, M. and {Cavalier}, F. and {Cavalieri}, R. and {Cella}, G. and {Cepeda}, C.~B. and {Cerd{\'a}-Dur{\'a}n}, P. and {Cerretani}, G. and {Cesarini}, E. and {Chamberlin}, S.~J. and {Chan}, M. and {Chao}, S. and {Charlton}, P. and {Chase}, E. and {Chassande-Mottin}, E. and {Chatterjee}, D. and {Chatziioannou}, K. and {Cheeseboro}, B.~D. and {Chen}, H.~Y. and {Chen}, X. and {Chen}, Y. and {Cheng}, H.-P. and {Chia}, H. and {Chincarini}, A. and {Chiummo}, A. and {Chmiel}, T. and {Cho}, H.~S. and {Cho}, M. and {Chow}, J.~H. and {Christensen}, N. and {Chu}, Q. and {Chua}, A.~J.~K. and {Chua}, S.},
        title = "{GW170817: Observation of Gravitational Waves from a Binary Neutron Star Inspiral}",
      journal = {Phys. Rev. Lett.},
         year = 2017,
        month = oct,
       volume = {119},
       number = {16},
          eid = {161101},
        pages = {161101},
          doi = {10.1103/PhysRevLett.119.161101},
archivePrefix = {arXiv},
       eprint = {1710.05832},
 primaryClass = {gr-qc},
       adsurl = {https://ui.adsabs.harvard.edu/abs/2017PhRvL.119p1101A}
}

@ARTICLE{2017ApJ...848L..12A,
       author = {{Abbott}, B.~P. and {Abbott}, R. and {Abbott}, T.~D. and {Acernese}, F. and {Ackley}, K. and {Adams}, C. and {Adams}, T. and {Addesso}, P. and {Adhikari}, R.~X. and {Adya}, V.~B. and {Affeldt}, C. and {Afrough}, M. and {Agarwal}, B. and {Agathos}, M. and {Agatsuma}, K. and {Aggarwal}, N. and {Aguiar}, O.~D. and {Aiello}, L. and {Ain}, A. and {Ajith}, P. and {Allen}, B. and {Allen}, G. and {Allocca}, A. and {Altin}, P.~A. and {Amato}, A. and {Ananyeva}, A. and {Anderson}, S.~B. and {Anderson}, W.~G. and {Angelova}, S.~V. and {Antier}, S. and {Appert}, S. and {Arai}, K. and {Araya}, M.~C. and {Areeda}, J.~S. and {Arnaud}, N. and {Arun}, K.~G. and {Ascenzi}, S. and {Ashton}, G. and {Ast}, M. and {Aston}, S.~M. and {Astone}, P. and {Atallah}, D.~V. and {Aufmuth}, P. and {Aulbert}, C. and {AultONeal}, K. and {Austin}, C. and {Avila-Alvarez}, A. and {Babak}, S. and {Bacon}, P. and {Bader}, M.~K.~M. and {Bae}, S. and {Baker}, P.~T. and {Baldaccini}, F. and {Ballardin}, G. and {Ballmer}, S.~W. and {Banagiri}, S. and {Barayoga}, J.~C. and {Barclay}, S.~E. and {Barish}, B.~C. and {Barker}, D. and {Barkett}, K. and {Barone}, F. and {Barr}, B. and {Barsotti}, L. and {Barsuglia}, M. and {Barta}, D. and {Barthelmy}, S.~D. and {Bartlett}, J. and {Bartos}, I. and {Bassiri}, R. and {Basti}, A. and {Batch}, J.~C. and {Bawaj}, M. and {Bayley}, J.~C. and {Bazzan}, M. and {B{\'e}csy}, B. and {Beer}, C. and {Bejger}, M. and {Belahcene}, I. and {Bell}, A.~S. and {Berger}, B.~K. and {Bergmann}, G. and {Bero}, J.~J. and {Berry}, C.~P.~L. and {Bersanetti}, D. and {Bertolini}, A. and {Betzwieser}, J. and {Bhagwat}, S. and {Bhandare}, R. and {Bilenko}, I.~A. and {Billingsley}, G. and {Billman}, C.~R. and {Birch}, J. and {Birney}, R. and {Birnholtz}, O. and {Biscans}, S. and {Biscoveanu}, S. and {Bisht}, A. and {Bitossi}, M. and {Biwer}, C. and {Bizouard}, M.~A. and {Blackburn}, J.~K. and {Blackman}, J. and {Blair}, C.~D. and {Blair}, D.~G. and {Blair}, R.~M. and {Bloemen}, S. and {Bock}, O. and {Bode}, N. and {Boer}, M. and {Bogaert}, G. and {Bohe}, A. and {Bondu}, F. and {Bonilla}, E. and {Bonnand}, R. and {Boom}, B.~A. and {Bork}, R. and {Boschi}, V. and {Bose}, S. and {Bossie}, K. and {Bouffanais}, Y. and {Bozzi}, A. and {Bradaschia}, C. and {Brady}, P.~R. and {Branchesi}, M. and {Brau}, J.~E. and {Briant}, T. and {Brillet}, A. and {Brinkmann}, M. and {Brisson}, V. and {Brockill}, P. and {Broida}, J.~E. and {Brooks}, A.~F. and {Brown}, D.~A. and {Brown}, D.~D. and {Brunett}, S. and {Buchanan}, C.~C. and {Buikema}, A. and {Bulik}, T. and {Bulten}, H.~J. and {Buonanno}, A. and {Buskulic}, D. and {Buy}, C. and {Byer}, R.~L. and {Cabero}, M. and {Cadonati}, L. and {Cagnoli}, G. and {Cahillane}, C. and {Calder{\'o}n Bustillo}, J. and {Callister}, T.~A. and {Calloni}, E. and {Camp}, J.~B. and {Canepa}, M. and {Canizares}, P. and {Cannon}, K.~C. and {Cao}, H. and {Cao}, J. and {Capano}, C.~D. and {Capocasa}, E. and {Carbognani}, F. and {Caride}, S. and {Carney}, M.~F. and {Casanueva Diaz}, J. and {Casentini}, C. and {Caudill}, S. and {Cavagli{\`a}}, M. and {Cavalier}, F. and {Cavalieri}, R. and {Cella}, G. and {Cepeda}, C.~B. and {Cerd{\'a}-Dur{\'a}n}, P. and {Cerretani}, G. and {Cesarini}, E. and {Chamberlin}, S.~J. and {Chan}, M. and {Chao}, S. and {Charlton}, P. and {Chase}, E. and {Chassande-Mottin}, E. and {Chatterjee}, D. and {Chatziioannou}, K. and {Cheeseboro}, B.~D. and {Chen}, H.~Y. and {Chen}, X. and {Chen}, Y. and {Cheng}, H.-P. and {Chia}, H. and {Chincarini}, A. and {Chiummo}, A. and {Chmiel}, T. and {Cho}, H.~S. and {Cho}, M. and {Chow}, J.~H. and {Christensen}, N. and {Chu}, Q. and {Chua}, A.~J.~K. and {Chua}, S. and {Chung}, A.~K.~W. and {Chung}, S. and {Ciani}, G.},
        title = "{Multi-messenger Observations of a Binary Neutron Star Merger}",
      journal = {Astrophys. J. Lett.},
         year = 2017,
        month = oct,
       volume = {848},
       number = {2},
          eid = {L12},
        pages = {L12},
          doi = {10.3847/2041-8213/aa91c9},
archivePrefix = {arXiv},
       eprint = {1710.05833},
 primaryClass = {astro-ph.HE},
       adsurl = {https://ui.adsabs.harvard.edu/abs/2017ApJ...848L..12A}
}

@ARTICLE{1999A&A...341..499R,
       author = {{Rosswog}, S. and {Liebend{\"o}rfer}, M. and {Thielemann}, F.-K. and {Davies}, M.~B. and {Benz}, W. and {Piran}, T.},
        title = "{Mass ejection in neutron star mergers}",
      journal = {Astron. Astrophys.},
         year = 1999,
        month = jan,
       volume = {341},
        pages = {499-526},
          doi = {10.48550/arXiv.astro-ph/9811367},
archivePrefix = {arXiv},
       eprint = {astro-ph/9811367},
 primaryClass = {astro-ph},
       adsurl = {https://ui.adsabs.harvard.edu/abs/1999A&A...341..499R}
}

@article{Goriely_2011,
doi = {10.1088/2041-8205/738/2/L32},
url = {https://doi.org/10.1088/2041-8205/738/2/L32},
year = {2011},
month = {aug},
publisher = {The American Astronomical Society},
volume = {738},
number = {2},
pages = {L32},
author = {Goriely, Stephane and Bauswein, Andreas and Janka, Hans-Thomas},
title = {r-PROCESS NUCLEOSYNTHESIS IN DYNAMICALLY EJECTED MATTER OF NEUTRON STAR MERGERS},
journal = {Astrophys. J. Lett.}
}

@article{Wanajo_2014,
doi = {10.1088/2041-8205/789/2/L39},
url = {https://doi.org/10.1088/2041-8205/789/2/L39},
year = {2014},
month = {jun},
publisher = {The American Astronomical Society},
volume = {789},
number = {2},
pages = {L39},
author = {Wanajo, Shinya and Sekiguchi, Yuichiro and Nishimura, Nobuya and Kiuchi, Kenta and Kyutoku, Koutarou and Shibata, Masaru},
title = {PRODUCTION OF ALL THE r-PROCESS NUCLIDES IN THE DYNAMICAL EJECTA OF NEUTRON STAR MERGERS},
journal = {Astrophys. J. Lett.}
}

@article{10.1093/mnras/stw2156,
    author = {Wu, Meng-Ru and Fernández, Rodrigo and Martínez-Pinedo, Gabriel and Metzger, Brian D.},
    title = {Production of the entire range of r-process nuclides by black hole accretion disc outflows from neutron star mergers},
    journal = {Mon. Not. R. Astron. Soc.},
    volume = {463},
    number = {3},
    pages = {2323-2334},
    year = {2016},
    month = {08},
    issn = {0035-8711},
    doi = {10.1093/mnras/stw2156},
    url = {https://doi.org/10.1093/mnras/stw2156},
    eprint = {https://academic.oup.com/mnras/article-pdf/463/3/2323/18240271/stw2156.pdf},
}

@ARTICLE{2017Natur.551...80K,
       author = {{Kasen}, Daniel and {Metzger}, Brian and {Barnes}, Jennifer and {Quataert}, Eliot and {Ramirez-Ruiz}, Enrico},
        title = "{Origin of the heavy elements in binary neutron-star mergers from a gravitational-wave event}",
      journal = {Nat.},
         year = 2017,
        month = nov,
       volume = {551},
       number = {7678},
        pages = {80-84},
          doi = {10.1038/nature24453},
archivePrefix = {arXiv},
       eprint = {1710.05463},
 primaryClass = {astro-ph.HE},
       adsurl = {https://ui.adsabs.harvard.edu/abs/2017Natur.551...80K}
}

@ARTICLE{2019Natur.574..497W,
       author = {{Watson}, Darach and {Hansen}, Camilla J. and {Selsing}, Jonatan and {Koch}, Andreas and {Malesani}, Daniele B. and {Andersen}, Anja C. and {Fynbo}, Johan P.~U. and {Arcones}, Almudena and {Bauswein}, Andreas and {Covino}, Stefano and {Grado}, Aniello and {Heintz}, Kasper E. and {Hunt}, Leslie and {Kouveliotou}, Chryssa and {Leloudas}, Giorgos and {Levan}, Andrew J. and {Mazzali}, Paolo and {Pian}, Elena},
        title = "{Identification of strontium in the merger of two neutron stars}",
      journal = {Nat.},
         year = 2019,
        month = oct,
       volume = {574},
       number = {7779},
        pages = {497-500},
          doi = {10.1038/s41586-019-1676-3},
archivePrefix = {arXiv},
       eprint = {1910.10510},
 primaryClass = {astro-ph.HE},
       adsurl = {https://ui.adsabs.harvard.edu/abs/2019Natur.574..497W}
}

@ARTICLE{2022MNRAS.515..631G,
       author = {{Gillanders}, J.~H. and {Smartt}, S.~J. and {Sim}, S.~A. and {Bauswein}, A. and {Goriely}, S.},
        title = "{Modelling the spectra of the kilonova AT2017gfo - I. The photospheric epochs}",
      journal = {Mon. Not. R. Astron. Soc.},
         year = 2022,
        month = sep,
       volume = {515},
       number = {1},
        pages = {631-651},
          doi = {10.1093/mnras/stac1258},
archivePrefix = {arXiv},
       eprint = {2202.01786},
 primaryClass = {astro-ph.HE},
       adsurl = {https://ui.adsabs.harvard.edu/abs/2022MNRAS.515..631G}
}

@article{osti_6189038,
  author       = {Sandulescu, A and Poenaru, D N and Greiner, W},
  title        = {New type of decay of heavy nuclei intermediate between fission and. cap alpha. decay},
  url          = {https://www.osti.gov/biblio/6189038},
  journal      = {Sov. J. Particles Nucl. (Engl. Transl.); (United States)},
  issn         = {ISSN SJPNA},
  volume       = {11:6},
  place        = {United States},
  year         = {1980},
  month        = {11}}

@article{rose_new_1984,
	title = {A new kind of natural radioactivity},
	volume = {307},
	issn = {1476-4687},
	url = {https://doi.org/10.1038/307245a0},
	doi = {10.1038/307245a0},
	number = {5948},
	journal = {Nat.},
	author = {Rose, H. J. and Jones, G. A.},
	month = jan,
	year = {1984},
	pages = {245--247},
}

@article{Poenaru_1984,
doi = {10.1088/0305-4616/10/8/004},
url = {https://doi.org/10.1088/0305-4616/10/8/004},
year = {1984},
month = {aug},
publisher = {},
volume = {10},
number = {8},
pages = {L183},
author = {D N Poenaru and M Ivascu and A Sandulescu and W Greiner},
title = {Spontaneous emission of heavy clusters},
journal = {J. Phys. G: Nucl. Phys.},

}

@article{JAIN2023122597,
title = {Cluster radioactivity in trans-lead region: A systematic study with modified empirical formulas},
journal = {Nucl. Phys. A},
volume = {1031},
pages = {122597},
year = {2023},
issn = {0375-9474},
doi = {https://doi.org/10.1016/j.nuclphysa.2022.122597},
url = {https://www.sciencedirect.com/science/article/pii/S0375947422002214},
author = {A. Jain and P.K. Sharma and S.K. Jain and J.K. Deegwal and G. Saxena}
}

@article{bonetti_guglielmetti_2007, title={Cluster radioactivity: an overview after twenty years}, volume={59}, ISSN={1221-1451}, abstractNote={The present status of experimental research in cluster radioactivity is reviewed with emphasis on results obtained in the last few years. Various theoretical approaches are briefly discussed and compared with recently obtained experimental results on 34Si and 22Ne clusters. Experiments in progress on 238U and 223Ac are described, and open problems in the experimental and theoretical research are outlined. (authors) Key words: cluster radioactivity, fission models, alpha decay models}, number={2}, journal={Rom. Rep. Phys.}, author={Bonetti, Roberto and Guglielmetti, Alessandra}, year={2007}, pages={p. 301–310} }

@article{ROYER2001182,
title = {Light nucleus emission within a generalized liquid-drop model and quasimolecular shapes},
journal = {Nucl. Phys. A},
volume = {683},
number = {1},
pages = {182-206},
year = {2001},
issn = {0375-9474},
doi = {https://doi.org/10.1016/S0375-9474(00)00454-1},
url = {https://www.sciencedirect.com/science/article/pii/S0375947400004541},
author = {G. Royer and R. Moustabchir}
}

@article{BONETTI199332,
title = {Nuclear structure effects in the exotic decay of 225{Ac} via {14C} emission},
journal = {Nucl. Phys. A},
volume = {562},
number = {1},
pages = {32-40},
year = {1993},
issn = {0375-9474},
doi = {https://doi.org/10.1016/0375-9474(93)90030-2},
url = {https://www.sciencedirect.com/science/article/pii/0375947493900302},
author = {R. Bonetti and C. Chiesa and A. Guglielmetti and R. Matheoud and C. Migliorino and A.L. Pasinetti and H.L. Ravn}
}

@article{warda2011,
  title = {Microscopic description of cluster radioactivity in actinide nuclei},
  author = {Warda, M. and Robledo, L. M.},
  journal = {Phys. Rev. C},
  volume = {84},
  issue = {4},
  pages = {044608},
  numpages = {17},
  year = {2011},
  month = {Oct},
  publisher = {American Physical Society},
  doi = {10.1103/PhysRevC.84.044608},
  url = {https://link.aps.org/doi/10.1103/PhysRevC.84.044608}
}

@article{PhysRevC.111.065803,
  title = {New upper limits for $\ensuremath{\beta}$-delayed fission probabilities of $^{230,232}\mathrm{Fr}$ and $^{230,232,234}\mathrm{Ac}$},
  author = {Bara, S. and Algora, A. and Andel, B. and Andreyev, A. N. and Antalic, S. and Bark, R. A. and Borge, M. J. G. and Camaiani, A. and Cocolios, T. E. and Cubiss, J. G. and De Witte, H. and Fajardo-Zambrano, C. M. and Favier, Z. and Fraile, L. M. and Fynbo, H. O. U. and Goriely, S. and Grzywacz, R. and Heines, M. and Ivandikov, F. and Johnson, J. D. and Jones, P. M. and Judson, D. S. and Klimo, J. and Korgul, A. and Labiche, M. and Lica, R. and Madurga, M. and Marginean, N. and Mihai, C. and Mi\v{s}t, J. and N\'acher, E. and Neacsu, C. and Orce, J. N. and Page, C. A. A. and Page, R. D. and Pakarinen, J. and Papadakis, P. and Perea, A. and Piersa-Si\l{}kowska, M. and Podoly\'ak, Zs. and Raabe, R. and Ryssens, W. and S\'anchez-Fern\'andez, A. and Sitar\v{c}\'{\i}k, A. and Tengblad, O. and Ud\'{\i}as, J. M. and Van Den Bergh, V. and Van Duppen, P. and Warr, N. and Youssef, A. and Yue, Z.},
  journal = {Phys. Rev. C},
  volume = {111},
  issue = {6},
  pages = {065803},
  numpages = {11},
  year = {2025},
  month = {Jun},
  publisher = {American Physical Society},
  doi = {10.1103/PhysRevC.111.065803},
  url = {https://link.aps.org/doi/10.1103/PhysRevC.111.065803}
}

@article{10.1093/mnras/stab3393,
    author = {Kullmann, I and Goriely, S and Just, O and Ardevol-Pulpillo, R and Bauswein, A and Janka, H-T},
    title = {Dynamical ejecta of neutron star mergers with nucleonic weak processes {I}: nucleosynthesis},
    journal = {Monthly Notices of the Royal Astronomical Society},
    volume = {510},
    number = {2},
    pages = {2804-2819},
    year = {2022},
    month = {02},
    issn = {0035-8711},
    doi = {10.1093/mnras/stab3393},
    url = {https://doi.org/10.1093/mnras/stab3393},
}

@article{goriely_fundamental_2015,
	title = {The fundamental role of fission during r-process nucleosynthesis in neutron star mergers},
	volume = {51},
	issn = {1434-601X},
	url = {https://doi.org/10.1140/epja/i2015-15022-3},
	doi = {10.1140/epja/i2015-15022-3},
	number = {2},
	journal = {Eur. Phys. J. A},
	author = {Goriely, S.},
	month = feb,
	year = {2015},
	pages = {22},
}

@article{SAXENA2024122867,
title = {Theoretical investigation of heavy cluster decay from {Z}=118 and 120 isotopes: A search for an empirical formula in superheavy region},
journal = {Nucl. Phys. A},
volume = {1046},
pages = {122867},
year = {2024},
issn = {0375-9474},
doi = {https://doi.org/10.1016/j.nuclphysa.2024.122867},
url = {https://www.sciencedirect.com/science/article/pii/S0375947424000496},
author = {G. Saxena and Dashty T. Akrawy and Ali H. Ahmed and Mamta Aggarwal}
}

@article{Just_2023,
doi = {10.3847/2041-8213/acdad2},
url = {https://doi.org/10.3847/2041-8213/acdad2},
year = {2023},
month = {jul},
publisher = {The American Astronomical Society},
volume = {951},
number = {1},
pages = {L12},
author = {Just, O. and Vijayan, V. and Xiong, Z. and Goriely, S. and Soultanis, T. and Bauswein, A. and Guilet, J. and Janka, H.-Th. and Martínez-Pinedo, G.},
title = {End-to-end Kilonova Models of Neutron Star Mergers with Delayed Black Hole Formation},
journal = {Astrophys. J. Lett.}
}

@ARTICLE{1957RvMP...29..547B,
       author = {{Burbidge}, E. Margaret and {Burbidge}, G.~R. and {Fowler}, William A. and {Hoyle}, F.},
        title = "{Synthesis of the Elements in Stars}",
      journal = {Reviews of Modern Physics},
         year = 1957,
        month = oct,
       volume = {29},
       number = {4},
        pages = {547-650},
          doi = {10.1103/RevModPhys.29.547},
       adsurl = {https://ui.adsabs.harvard.edu/abs/1957RvMP...29..547B}
}

@article{PhysRevC.70.017301,
  title = {New semiempirical formula for exotic cluster decay},
  author = {Balasubramaniam, M. and Kumarasamy, S. and Arunachalam, N. and Gupta, Raj K.},
  journal = {Phys. Rev. C},
  volume = {70},
  issue = {1},
  pages = {017301},
  numpages = {4},
  year = {2004},
  month = {Jul},
  publisher = {American Physical Society},
  doi = {10.1103/PhysRevC.70.017301},
  url = {https://link.aps.org/doi/10.1103/PhysRevC.70.017301}
}

@article{BROWNE20111115,
title = {Nuclear Data Sheets for {A} = 220},
journal = {Nuclear Data Sheets},
volume = {112},
number = {4},
pages = {1115-1161},
year = {2011},
issn = {0090-3752},
doi = {https://doi.org/10.1016/j.nds.2011.03.002},
url = {https://www.sciencedirect.com/science/article/pii/S0090375211000214},
author = {E. Browne and J.K. Tuli}
}

@phdthesis{ghysphd,
title = "beta-delayed fission in proton-rich nuclei in the lead region",
school = "KU Leuven",
author = "Lars Ghys",
year = "2015",
month = sep,
language = "English",
publisher = "KUL - Katholieke Universiteit Leuven",
}

@article{MORSE2024259,
title = {Nuclear Structure and Decay Data for {A}=230},
journal = {Nuclear Data Sheets},
volume = {197},
pages = {259-371},
year = {2024},
issn = {0090-3752},
doi = {https://doi.org/10.1016/j.nds.2024.08.002},
url = {https://www.sciencedirect.com/science/article/pii/S0090375224000528},
author = {C. Morse}
}

@article{HOFFMAN200917,
title = {Evidence for a doubly magic 24{O}},
journal = {Physics Letters B},
volume = {672},
number = {1},
pages = {17-21},
year = {2009},
issn = {0370-2693},
doi = {https://doi.org/10.1016/j.physletb.2008.12.066},
url = {https://www.sciencedirect.com/science/article/pii/S0370269309000069},
author = {C.R. Hoffman and T. Baumann and D. Bazin and J. Brown and G. Christian and D.H. Denby and P.A. DeYoung and J.E. Finck and N. Frank and J. Hinnefeld and S. Mosby and W.A. Peters and W.F. Rogers and A. Schiller and A. Spyrou and M.J. Scott and S.L. Tabor and M. Thoennessen and P. Voss}
}

@misc{224Ra,
    author = {B. Singh and S. Singh},
    title = {224Ra Adopted levels, Gammas},
    note = {From ENSDF database as of July 7, 2026.
Version available at http://www.nndc.bnl.gov/ensarchivals/}
}

@article{sridhar_atlas_2020,
	title = {Atlas of cluster radioactivity in actinide nuclei},
	volume = {135},
	issn = {2190-5444},
	url = {https://doi.org/10.1140/epjp/s13360-020-00302-1},
	doi = {10.1140/epjp/s13360-020-00302-1},
	number = {3},
	journal = {The European Physical Journal Plus},
	author = {Sridhar, G. R. and Manjunatha, H. C. and Sowmya, N. and Gupta, P. S. Damodara and Ramalingam, H. B.},
	month = mar,
	year = {2020},
	pages = {291},
}

@article{PhysRevLett.103.072501,
  title = {Universal Decay Law in Charged-Particle Emission and Exotic Cluster Radioactivity},
  author = {Qi, C. and Xu, F. R. and Liotta, R. J. and Wyss, R.},
  journal = {Phys. Rev. Lett.},
  volume = {103},
  issue = {7},
  pages = {072501},
  numpages = {4},
  year = {2009},
  month = {Aug},
  publisher = {American Physical Society},
  doi = {10.1103/PhysRevLett.103.072501},
  url = {https://link.aps.org/doi/10.1103/PhysRevLett.103.072501}
}

@article{Qi_2023,
doi = {10.1088/1674-1137/accc78},
url = {https://doi.org/10.1088/1674-1137/accc78},
year = {2023},
month = {jun},
publisher = {Chinese Physical Society and the Institute of High Energy Physics of the Chinese Academy of Sciences and the Institute of Modern Physics of the Chinese Academy of Sciences and IOP Publishing Ltd},
volume = {47},
number = {6},
pages = {064107},
author = {Qi, Lin-Jing and Zhang, Dong-Meng and Luo, Song and He, Biao and Wu, Xi-Jun and Chen, Xun and Li, Xiao-Hua},
title = {{New Geiger-Nuttall} law for cluster radioactivity half-lives},
journal = {Chinese Physics C}
}

@article{Wang_2021,
doi = {10.1088/1674-1137/abddaf},
url = {https://doi.org/10.1088/1674-1137/abddaf},
year = {2021},
month = {mar},
publisher = {Chinese Physical Society and the Institute of High Energy Physics of the Chinese Academy of Sciences and the Institute of Modern Physics of the Chinese Academy of Sciences and IOP Publishing Ltd},
volume = {45},
number = {3},
pages = {030003},
author = {Wang, Meng and Huang, W.J. and Kondev, F.G. and Audi, G. and Naimi, S.},
title = {The {AME} 2020 atomic mass evaluation ({II}). Tables, graphs and references},
journal = {Chinese Physics C}
}

@article{grams_skyrme-hartree-fock-bogoliubov_2023,
	title = {Skyrme-{Hartree}-{Fock}-{Bogoliubov} mass models on a {3D} mesh: {III}. {From} atomic nuclei to neutron stars},
	volume = {59},
	issn = {1434-601X},
	url = {https://doi.org/10.1140/epja/s10050-023-01158-6},
	doi = {10.1140/epja/s10050-023-01158-6},
	number = {11},
	journal = {The European Physical Journal A},
	author = {Grams, Guilherme and Ryssens, Wouter and Scamps, Guillaume and Goriely, Stephane and Chamel, Nicolas},
	month = nov,
	year = {2023},
	pages = {270},
}

@article{SINGH201970,
title = {Nuclear Data Sheets for {A}=266,270,274,278,282,286,290,294,298},
journal = {Nuclear Data Sheets},
volume = {156},
pages = {70-147},
year = {2019},
issn = {0090-3752},
doi = {https://doi.org/10.1016/j.nds.2019.02.004},
url = {https://www.sciencedirect.com/science/article/pii/S0090375219300171},
author = {Balraj Singh}
}

@article{PhysRevC.32.572,
  title = {Atomic nuclei decay modes by spontaneous emission of heavy ions},
  author = {Poenaru, D. N. and Iva{\c{s}}cu, M. and Sandulescu, A. and Greiner, Walter},
  journal = {Phys. Rev. C},
  volume = {32},
  issue = {2},
  pages = {572--581},
  numpages = {0},
  year = {1985},
  month = {Aug},
  publisher = {American Physical Society},
  doi = {10.1103/PhysRevC.32.572},
  url = {https://link.aps.org/doi/10.1103/PhysRevC.32.572}
}

@misc{bara_2024_14358715,
  author       = {Bara, Silvia and
                  Andel, Boris and
                  Andreyev, Andrei N. and
                  Antalic, Stanislav and
                  Camaiani, Alberto and
                  Cocolios, Thomas E. and
                  Cubiss, James G. and
                  De Witte, Hilde and
                  Fajardo-Zambrano, Carlos M. and
                  Favier, Zoe and
                  Goriely, Stéphane and
                  Heines, Michael and
                  Ivandikov, Fedor and
                  Johnson, Jake D. and
                  Klimo, Jozef and
                  Lica, Razvan and
                  Mist, Jozef and
                  Page, Chris and
                  Raabe, Riccardo and
                  Ryssens, Wouter and
                  Sitarcik, Adam and
                  Van Den Bergh, Viktor and
                  Van Duppen, Piet and
                  Youssef, Ahmed and
                  Yue, Zixuan},
  title        = {Open Dataset for publication: New upper limits
                   for beta-delayed fission probabilities of
                   230,232Fr and 230,232,234Ac - LOI216},
  month        = dec,
  year         = 2024,
  publisher    = {Zenodo},
  doi          = {10.5281/zenodo.14358715},
}

@article{b7q7-925c,
  title = {Microscopic description of cluster radioactivity fission valleys along isotopic and isotonic chains},
  author = {Warda, M. and Zdeb, A. and Rodr\'{\i}guez-Guzm\'an, R.},
  journal = {Phys. Rev. C},
  volume = {113},
  issue = {3},
  pages = {034619},
  numpages = {11},
  year = {2026},
  month = {Mar},
  publisher = {American Physical Society},
  doi = {10.1103/b7q7-925c},
  url = {https://link.aps.org/doi/10.1103/b7q7-925c}
}

@article{Koura01082002,
author = {Hiroyuki Koura and Takahiro Tachibana and Tadashi Yoshida},
title = {Estimation of Alpha-decay Half-lives and Fission Barriers from the Viewpoint of a New Mass Formula},
journal = {Journal of Nuclear Science and Technology},
volume = {39},
number = {sup2},
pages = {774--777},
year = {2002},
publisher = {Taylor \& Francis},
doi = {10.1080/00223131.2002.10875212},
URL = {https://doi.org/10.1080/00223131.2002.10875212},
eprint = {https://doi.org/10.1080/00223131.2002.10875212}
}

@article{PhysRevC.93.025805,
  title = {Large-scale evaluation of $\ensuremath{\beta}$-decay rates of $r$-process nuclei with the inclusion of first-forbidden transitions},
  author = {Marketin, T. and Huther, L. and Mart\'{\i}nez-Pinedo, G.},
  journal = {Phys. Rev. C},
  volume = {93},
  issue = {2},
  pages = {025805},
  numpages = {17},
  year = {2016},
  month = {Feb},
  publisher = {American Physical Society},
  doi = {10.1103/PhysRevC.93.025805},
  url = {https://link.aps.org/doi/10.1103/PhysRevC.93.025805}
}
\end{document}